\documentclass[12pt]{article}
\usepackage{ctex}
\usepackage{amssymb,amsmath,color,amsmath,bm}
\usepackage[colorlinks,linkcolor=blue]{hyperref}
\usepackage{geometry}
\usepackage{tabularx}
\usepackage{bm}
\usepackage{authblk}
\usepackage[square, comma, sort&compress, numbers]{natbib}

	\title{\bf The $[1,0]$-twisted generalized Reed-Solomon code}	
\author{\small Canze Zhu$^{1}$}
\author{\small Qunying Liao$^{2}$
	\thanks{{Corresponding author$^{2}$. E-mail: qunyingliao@sicnu.edu.cn;
			
			{\quad Contributing author$^{1}$. E-mail: ~canzezhu@163.com.}}	
	}}
\affil[]{\small (College of Mathematical Science, Sichuan Normal University, Chengdu Sichuan, 610066)}
\date{}
\newtheorem{theorem}{Theorem}[section]
\newtheorem{definition}{Definition}[section]
\newtheorem{lemma}{Lemma}[section]

\newtheorem{proposition}{Proposition}[section]
\newtheorem{corollary}{Corollary}[section]
\newtheorem{remark}{Remark}[section]

\catcode`@=11 \@addtoreset{equation}{section} \catcode`@=12
\begin{document}
	\maketitle
	{\bf Abstract.}
	{\small In this paper, we  not only give  the parity check  matrix of the $[1,0]$-twisted generalized Reed-Solomon (in short, TGRS) code, but also determine the weight distribution. Especially, we show that the $[1,0]$-TGRS code is not GRS or EGRS. Furthermore,  we present a  sufficient and necessary condition for any punctured code of the $[1,0]$-TGRS code to be self-orthogonal, and then construct several classes of self-dual or almost self-dual $[1,0]$-TGRS codes.  Finally, basing on these self-dual or almost self-dual  $[1,0]$-TGRS codes, we obtain some LCD $[1,0]$-TGRS codes.
		%self-orthogonal $[1,0]$-TGRS codes and  $[1,0]$-TGRS codes, some of these codes are self-dual or almost self-dual.
		%		 by using some properties for linear algebra methods, the parity check matrixs for twisted generalized Reed-Solomon codes with any given hook $h$ and twist $t$ are presented, and then a sufficient and necessary condition for that a twisted generalized Reed-Solomon code with $h\ge t$ to be self-dual is given. Furthermore, several classes of self-dual codes with small singleton defect are constructed based on twisted generalized Reed-Solomon codes.
	}\\
	
	
	\section{Introduction}
	
	An $[n,k,d]$ linear code $\mathcal{C}$ over $\mathbb{F}_q$ is a $k$-dimensional subspace of $\mathbb{F}_q^n$ with minimum (Hamming) distance $d$. If the
	parameters reach the Singleton bound, namely, $d=n-k+1$, then $\mathcal{C}$  is  maximum distance separable (in short, MDS). For the near-optimal case,  namely, $d=n-k$, 
	$\mathcal{C}$ is almost MDS (in short, AMDS). $\mathcal{C}$ is said to be near MDS (in short, NMDS) if both $\mathcal{C}$ and $\mathcal{C}^{\perp}$ are AMDS. Since MDS codes and NMDS codes are very important in coding theory and applications \cite{7,17,28,35,43}, the study of MDS codes or near MDS codes, including weight distributions, constructions, equivalence, self-dual properties, linear complementary dual (in short, LCD) properties, and so on, has attracted a lot of attention \cite{1,11,12,13,14,15,20,21,33,34,37}. Especially, generalized Reed-Solomon (in short, GRS) codes are a class of MDS codes. A lot of self-dual, almost self-dual or LCD MDS codes are constructed  based on GRS codes \cite{FX20,HZ21,10,19,25,39,40,LCD1,LCD2, LCD3}.

	In 2017,  inspired by the construction for twisted Gabidulin codes \cite{t26}, Beelen et al. firstly introduced  twisted Reed-Solomon (in short, TRS) codes, which is a generalization for Reed-Solomon codes, they also showed that TRS codes could be well decoded.  Different from GRS codes, they showed that a twisted generalized Reed-Solomon  (in short, TGRS) code is not necessarily MDS and presented a  sufficient and necessary condition for a TGRS code to be MDS  \cite{2}. Especially, the authors showed that most of TGRS MDS codes are not GRS when the code rate  is less than one half \cite{3,26,2022}. Later, by TGRS codes, Lavauzelle et al. presented an efficient key-recovery attack used in the McEliece cryptosystem \cite{24}. After that, Huang et al. not only  showed that the $[1,0]$-TGRS code is MDS or NMDS, but also constructed several classes of LCD $[1,0]$-TGRS codes \cite{16}.  Furthermore, they presented a sufficient and necessary condition for the $[1,k-1]$-TGRS code to be self-dual, and then constructed several classes of self-dual MDS or NMDS codes \cite{t0}. Recently, we generalized Huang et al's results, and obtained some self-dual TGRS codes with small Singleton defect \cite{ZC21}.  More relative results about the construction for self-orthogonal or LCD  TGRS codes can be seen in \cite{26,SD0,SD1,SD2}.

	In this paper, we focus on the $[1,0]$-twisted generalized Reed Solomon (in short, TGRS) code. This paper is organized as follows. In section 2, some basic notations and results about linear codes are given. In section 3,  the parity check  matrix and the weight distribution of the $[1,0]$-TGRS code are obtained, and then we show that the $[1,0]$-TGRS code is not GRS or EGRS. In section 4, a  sufficient and necessary condition for any punctured code of the $[1,0]$-TGRS code to be self-orthogonal is presented, and then several classes of self-dual, almost self-dual or LCD $[1,0]$-TGRS codes are constructed. In section 5, we conclude the whole paper.
%	In this paper,  the parity check matrix for a TGRS code with any given hook $h$ and twist $t$ is presented. And then
%	a sufficient and necessary condition for TGRS codes with $h\ge t$ to be self-dual is given. Furthermore, several classes of self-dual $m$-MDS codes with $m\le\min\{t,h+1\}$  are constructed. Especially, a sufficient and necessary condition for self-dual TGRS codes with dimension $k$, hook $k-1$ and twist $2$ to be MDS, NMDS or $2$-MDS is given, respectively. And then several classes of self-dual $m$-MDS $(m=0,1~\text{or}~2)$ codes are obtained. We extends some results in \cite{t0}.
	
%	This paper is organized as follows. In section 2, some basic notations and results about TGRS codes are given. In section 3, the parity check matrixs for TGRS codes are obtained. In section 4, a sufficient and necessary condition for TGRS codes with $h\ge t$ to be self-dual is presented, and several classes of self-dual $m$-MDS codes with small $m$ are constructed. In section 5, we conclude the whole paper.
	\section{Preliminaries}
	
	Throughout this paper, we fix some notations as follows for convenience.\\
	
	$\bullet$  $q$ is a power of  a prime.\\	
	
	$\bullet$  $\mathbb{F}_q$ is the finite field with $q$ elements.\\
	
	$\bullet$ $\mathbb{F}_q^{*}=\mathbb{F}_q\backslash\{0\}$.\\
	%	$\bullet$  $\mathrm{C}_{0}^{2,q}=\{\alpha\in\mathbb{F}_{q}^{*}~|~\alpha~\text{ is a square element}\}$.\\
	
	$\bullet$  $\mathbb{F}_q[x]$ is the polynomial ring over $\mathbb{F}_q$.\\
	
	$\bullet$  $k$ and $n$ are both positive integers with $3\le k<n$.\\
	
	$\bullet$  $\boldsymbol{1}=(1,\ldots,1)\in\mathbb{F}_q^{n}$, ~$\boldsymbol{0}=(0,\ldots,0)\in\mathbb{F}_q^{n}$.\\
	
	$\bullet$  ${\bf O}_{k\times k}$ denotes the $k\times k$ zero matrix over $\mathbb{F}_q$.\\
	
	$\bullet$  For any $\boldsymbol{\alpha}=(\alpha_1,\ldots,\alpha_n)\in\mathbb{F}_q^n$, denote
	$$\boldsymbol{\alpha}^i=\begin{cases}
	(1,\ldots,1),&~~\text{if~}i=0;\\
	(\alpha_1^i,\ldots,\alpha_n^{i}),&~~\text{if~}i\in\mathbb{N}^{+},
	\end{cases}$$ and
	$$~A_{\boldsymbol{\alpha}}=\{\alpha_i~|~i=1,\ldots,n\},~~~
	P_{\boldsymbol{\alpha}}=\prod\limits_{\alpha\in A_{\boldsymbol{\alpha}}}\alpha,~~~ P_{\boldsymbol{\alpha},i}=(-1)^{n}\prod\limits_{\alpha\in A_{\boldsymbol{\alpha}}\backslash\{\alpha_i\}}\alpha.$$
	
	%(\boldsymbol{\alpha},a)=(\alpha_1,\ldots,\alpha_n,a)~~(\forall a\in\mathbb{F}_{q})~,
	
	$\bullet$ $S_n$ is the permutation group with order $n$.\\
	
	$\bullet$ $1_{\pi}$ is the identity in  $S_n$.\\
	%	$\bullet$ $r$ and $m$ are both positive integers with $r\mid m$ and $2r\le m$.\\
	%	$\bullet$ $\mathrm{Tr}_r^{m}$ is the trace map from $\mathbb{F}_{q^m}$ to $\mathbb{F}_{q^r}$. 	Namely, $$\mathrm{Tr}_r^{m}(x)=x^{p^{m-r}}+x^{p^{m-2r}}+\cdots+x.$$
	%	$\bullet$ For $\boldsymbol{s}=(\epsilon_1,\epsilon_2,\delta_1,\delta_2)\in\mathbb{F}_{q}^{4}$, denote $$D_{\boldsymbol{s},-}=\epsilon_1\delta_2-\epsilon_2\delta_1\text{ ~and~}  D_{\boldsymbol{s},+}=\epsilon_1\delta_2+\epsilon_2\delta_1. $$
	%	$\bullet$ For $\boldsymbol{\alpha}=(\alpha_1,\ldots,\alpha_{n})\in\mathbb{F}_{q}^{n}$ with $\alpha_i\neq \alpha_j$ $(i\neq j)$, denote $$A_{\boldsymbol{\alpha}}=\{\alpha_1,\ldots,\alpha_n\}.$$
	
	In this section, we  review some basic notations and knowledge about GRS codes, EGRS codes, $[1,0]$-TGRS codes, Schur product, punctured codes, self orthogonal codes, LCD codes,  NMDS codes, respectively.

	\subsection{GRS, EGRS and $[1,0]$-TGRS codes}
	\subsubsection{GRS and EGRS codes}
The definitions of the GRS code and the EGRS code are given in the following, respectively.
\begin{definition}[\cite{17}]\label{d0} 
	Let $\boldsymbol{\alpha}=(\alpha_1,\ldots,\alpha_n)\in\mathbb{F}_q^n$ with $\alpha_i\neq \alpha_j$ $(i\neq j)$ and
	$\boldsymbol{v} = (v_1,\ldots,v_{n})\in (\mathbb{F}_q^{*})^{n}$. Then
	the GRS code is defined as
	\begin{align*}
	\mathcal{GRS}_{k,n}(\boldsymbol{\alpha},\boldsymbol{v})=\{(v_1f({\alpha_{1}}),\ldots,v_nf(\alpha_{n}))~|~f(x)\in\mathbb{F}_q[x],~\deg f(x)\le k-1\}.
	\end{align*}	
	The EGRS code is defined as
	\begin{align*}
	\mathcal{GRS}_{k,n}(\boldsymbol{\alpha},\boldsymbol{v},\infty)=\{(v_1f({\alpha_{1}}),\ldots,v_nf(\alpha_{n}),f_{k-1})~|~f(x)\in\mathbb{F}_q[x],~\deg f(x)\le k-1\},
	\end{align*}	
	where  $f_{k-1}$ is the coefficient of $x^{k-1}$ in $f(x)$.
	
	Especially, $\mathcal{GRS}_{k,n}(\boldsymbol{\alpha},\boldsymbol{1})$ and $\mathcal{GRS}_{k,n}(\boldsymbol{\alpha},\boldsymbol{1},\infty)$ are called RS code and ERS code, respectively.
\end{definition}

The dual codes of the GRS code and the EGRS code are given in the following, respectively.
\begin{lemma}[\cite{19}]\label{drs}
	Let $\boldsymbol{u}=(u_1,\ldots,u_n)$ with $u_j=-\prod\limits_{i=1,i\neq j}^{n}(\alpha_j-\alpha_i)^{-1}$, then
	$$\big(\mathcal{GRS}_{k,n}(\boldsymbol{\alpha},\boldsymbol{1})\big)^{\perp}=\mathcal{GRS}_{n-k,n}(\boldsymbol{\alpha},\boldsymbol{u})$$ and
	$$\big(\mathcal{GRS}_{k,n}(\boldsymbol{\alpha},\boldsymbol{1},\infty)\big)^{\perp}=\mathcal{GRS}_{n+1-k,n}(\boldsymbol{\alpha},\boldsymbol{u},\infty).$$
\end{lemma}

	\subsubsection{$[1,0]$-TGRS codes}
	\begin{definition}[\cite{2}]\label{d23} 
		Let $t$, $h$ and $k$ be positive integers with $0\le h<k\le q$. For $\eta\in\mathbb{F}_q^{*}$, the set of $(k,t,h,\eta)$-twisted polynomials is defined as 
		\begin{align*}
			\mathcal{V}_{k,t,h,\eta}=\Big\{f(x)=\sum\limits_{i=0}^{k-1}a_ix^i+\eta a_hx^{k-1+t}~|~a_i\in\mathbb{F}_q~(i=0,\ldots,k-1)\Big\},
		\end{align*}
		which is a $k$-dimensional $\mathbb{F}_q$-linear subspace. $h$ and $t$ are called the hook and  the twist, respectively.
	\end{definition}

	From the twisted polynomials linear  space $\mathcal{V}_{k,1,0,\eta}$, the definition of the $[1,0]$-TGRS code is given as follows.
	\begin{definition}[\cite{2}]\label{d1} 
		Let $\eta\in\mathbb{F}_q^{*}$, $\boldsymbol{\alpha}=(\alpha_1,\ldots,\alpha_n)\in\mathbb{F}_q^n$ with $\alpha_i\neq \alpha_j$ $(i\neq j)$ and
		$\boldsymbol{v} = (v_1,\ldots,v_{n})\in (\mathbb{F}_q^{*})^{n}$. Then
		the $[1,0]$-TGRS code is defined as
		\begin{align*}
		\mathcal{C}_{k,n}(\boldsymbol{\alpha},\boldsymbol{v},\eta)=\{(v_1f({\alpha_{1}}),\ldots,v_nf(\alpha_{n}))~|~f(x)\in  \mathcal{V}_{k,1,0,\eta}\}.
		\end{align*}	
	 Especially,  $\mathcal{C}_{k,n}(\boldsymbol{\alpha},\boldsymbol{1},\eta)$ is called the $[1,0]$-TRS code.
	\end{definition}
%	\begin{remark} If $A_{\boldsymbol{\alpha}}=\mathbb{F}_q^{*}$, we denote $\mathcal{C}_{k,n}(\mathbb{F}_q^{*},\boldsymbol{v},\eta)=\mathcal{C}_{k,n}(\boldsymbol{\alpha},\boldsymbol{v},\eta)$.
%		If $A_{\boldsymbol{\alpha}}=\mathbb{F}_q$, we  denote $\mathcal{C}_{k,q}(\mathbb{F}_q,\boldsymbol{v},\eta)=\mathcal{C}_{k,n}(\boldsymbol{\alpha},\boldsymbol{v},\eta)$.
		% Ovbiously, any $[1,0]$-TGRS code can be seen as a punctured code for $\mathcal{C}_{k,q}(\mathbb{F}_q,\boldsymbol{v},\eta)$.
%	\end{remark}

	By Definition \ref{d23}, it is easy to see that the generator matrix of $\mathcal{C}_{k,n}(\boldsymbol{\alpha},\boldsymbol{v},\eta)$ is
		\begin{align}\label{G1}
	G_k=\left(\begin{matrix}
	&v_1(1+\eta\alpha_1^{k})~&v_2(1+\eta\alpha_2^{k})~&\ldots~&v_n(1+\eta\alpha_n^{k})~\\
	&v_1\alpha_1~&v_2\alpha_2~&\ldots~&v_n\alpha_{n}~\\
	&\vdots~&\vdots&~&\vdots~\\
	&v_1\alpha_1^{k-2}~&v_2\alpha_{2}^{k-2}~&\cdots~&v_n\alpha_{n}^{k-2}~\\
	&v_1\alpha_1^{k-1}~&v_2\alpha_2^{k-1}~&\ldots~&v_n\alpha_{n}^{k-1}~\\
	\end{matrix}\right).
	\end{align}
	
	For $P_{\boldsymbol{\alpha}}\neq 0$, Huang et al.  gave a parity check matrix for  $\mathcal{C}_{k,n}(\boldsymbol{\alpha},\boldsymbol{v},\eta)$ as follows.
	\begin{lemma}[Theorem $1$ \cite{16}]\label{DH}
		Let $a=\prod\limits_{i=1}^{n}\alpha_i\neq 0$ and $u_j=\prod\limits_{i=1, i\neq j}^{n}(\alpha_j-\alpha_i)^{-1}$ for any $j=1,\ldots,n$, then $\mathcal{C}_{k,n}(\boldsymbol{\alpha},\boldsymbol{v},\eta)$ has the parity check matrix
			\begin{align*}
		H^{*}_{n-k}=\left(\begin{matrix}
		&\frac{u_1}{v_1}~&\frac{u_2}{v_2}~&\ldots~&\frac{u_n}{v_n}~\\
		&\frac{u_1}{v_1}\alpha_1~&\frac{u_2}{v_2}\alpha_2~&\ldots~&\frac{u_n}{v_n}\alpha_n~\\
		&\vdots~&\vdots&~&\vdots~\\
		&\frac{u_1}{v_1}\big(\alpha_{1}^{n-k-1}-\eta \frac{a}{\alpha_1}\big)~&\frac{u_2}{v_2}\big(\alpha_{2}^{n-k-1}-\eta \frac{a}{\alpha_2}\big)~&\cdots~&\frac{u_n}{v_n}\big(\alpha_{n}^{n-k-1}-\eta\frac{a}{\alpha_n}\big)~\\
		\end{matrix}\right).
		\end{align*}
	\end{lemma}

	The following lemma shows that $\mathcal{C}_{k,n}(\boldsymbol{\alpha},\boldsymbol{v},\eta)$ is MDS or NMDS.
	\begin{lemma}[Lemma $1$ \cite{16}]\label{SNN}
		Let \begin{align}
			S_{k}(\boldsymbol{\alpha})=\Big\{(-1)^{k}\prod\limits_{i\in I}^{n}\alpha_i~|~I\subsetneq\{1,\ldots,n\},~|I|=k\Big\}, 
		\end{align}
		then the following two assertions hold.
		
		$(1)$ $\mathcal{C}_{k,n}(\boldsymbol{\alpha},\boldsymbol{v},\eta)$ is MDS if and only if $\eta^{-1}\in\mathbb{F}_{q}^{*}\backslash S_{k}(\boldsymbol{\alpha})$;
		
		$(2)$ $\mathcal{C}_{k,n}(\boldsymbol{\alpha},\boldsymbol{v},\eta)$ is NMDS if and only if $\eta^{-1}\in S_{k}(\boldsymbol{\alpha})$.
	\end{lemma}

	\begin{remark}\label{SNNT}
	In Theorem $13$ of \cite{2022},  if $q$ is an odd prime power, $3\le k\le\frac{q-1}{2}-2$ and $n>\frac{q+1}{2}$, the authors showed that $\mathcal{C}_{k,n}(\boldsymbol{\alpha},\boldsymbol{v},\eta)$ is not MDS. Then by Lemma \ref{SNN}, we can know that $\mathcal{C}_{k,n}(\boldsymbol{\alpha},\boldsymbol{v},\eta)$ is NMDS under the same assumptions.
	\end{remark}

\subsection{Some notations for linear codes}
\subsubsection{The Schur product}
The Schur product  is defined as follows.
\begin{definition}\label{SP}
	For $\mathbf{x}=(x_1,\ldots,x_n ), \mathbf{y}= (y_1 ,\dots, y_n)\in \mathbb{F}_q^{n}$,  the Schur product  between $\mathbf{x}$ and $\mathbf{y}$ is
	defined as $$\mathbf{x}\star\mathbf{y}:=(x_1y_1,\ldots,x_ny_n).$$  The Schur product  of two $q$-ary codes $\mathcal{C}_1$ and $\mathcal{C}_2$ with length $n$ is
	defined as
	\begin{align*}
	\mathcal{C}_1\star\mathcal{C}_2=\langle \mathbf{c}_1 \star\mathbf{c}_2~|~\mathbf{c}_1\in\mathcal{C}_1,\mathbf{c}_2\in\mathcal{C}_2\rangle.
	\end{align*}
	%	where $\langle S\rangle$ denotes the $\mathbb{F}_q$-linear subspace generated by the subset $S\subseteq \mathbb{F}_q^n$.
	Especially, for a code $\mathcal{C}$, we call $\mathcal{C}^2:=\mathcal{C}\star\mathcal{C}$ the Schur square of $\mathcal{C}$.
\end{definition} 
\begin{remark}\label{r1}For any linear codes  $\mathcal{C}_1$ and $\mathcal{C}_2$, if 
	$\mathcal{C}_1=\langle \boldsymbol{v}_1,\ldots,\boldsymbol{v}_{k_1}\rangle$ and   $\mathcal{C}_2=\langle \boldsymbol{w}_1,\ldots,\boldsymbol{w}_{k_2}\rangle$ with $\boldsymbol{v}_i,\boldsymbol{w}_j\in\mathbb{F}_q^{n}~(i=1,\ldots,k_1,~j=1,\ldots,k_2)$, then 
	\begin{align}\label{S1}
	\mathcal{C}_1\star\mathcal{C}_2=\langle \boldsymbol{v}_i\star\boldsymbol{w}_j~(i=1,\ldots,k_1,~j=1,\ldots,k_2)\rangle, 
	\end{align}
\end{remark}

By the definitions of the GRS code and the EGRS code, Lemma \ref{drs} and Remark \ref{r1}, we can get the following proposition directly.
\begin{proposition}\label{pr}	Let $\boldsymbol{u}=(u_1,\ldots,u_n)$, where $u_j=-\prod\limits_{i=1,i\neq j}^{n}(\alpha_j-\alpha_i)$ for any $j=1,\ldots,n$.
	
	$(1)$ If $k\le\frac{n}{2}$, then {\small \begin{align*}
		\mathcal{GRS}^{2}_{k,n}(\boldsymbol{\alpha},\boldsymbol{1})=\mathcal{GRS}_{2k-1,n}(\boldsymbol{\alpha},\boldsymbol{1}) \text{~~~and~~~}\mathcal{GRS}_{k,n-1}^{2}(\boldsymbol{\alpha},\boldsymbol{1},\infty)=\mathcal{GRS}_{2k-1,n-1}(\boldsymbol{\alpha},\boldsymbol{1},\infty).
		\end{align*}}
	
	$(2)$ If $n\ge  k\ge\frac{n}{2}-1$, then {\small\begin{align*}
		\big(\mathcal{GRS}_{k,n}^{\perp}(\boldsymbol{\alpha},\boldsymbol{1})\big)^2=\mathcal{GRS}_{2n-2k-1,n}(\boldsymbol{\alpha},\boldsymbol{u}^2)\text{~~~and~~~}\big(\mathcal{GRS}_{k,n}^{\perp}(\boldsymbol{\alpha},\boldsymbol{1},\infty)\big)^2=\mathcal{GRS}_{2n-2k-1,n-1}(\boldsymbol{\alpha},\boldsymbol{u}^2,\infty).
		\end{align*}}
	%For an $[n+1,k]$ GRS codes $\mathcal{C}_1$, or an $[n+1,k]$ EGRS codes $\mathcal{C}_2$, if $2k\le n+1$, then $\dim(\mathcal{C}_1^{2})=\dim(\mathcal{C}_2^{2})=2k-1$. 
\end{proposition}
\subsubsection{The equivalence and punctured codes for linear codes}
The definition of the equivalence for linear codes is given as follows.
\begin{definition}[\cite{28}]
	Let $\mathcal{C}_1$ and $\mathcal{C}_2$ be linear codes over $\mathbb{F}_{q}$ with length $n$. We say that $\mathcal{C}_1$ and $\mathcal{C}_2$ are equivalent if
	there exists a permutation $\pi\in S_n$ and $\boldsymbol{v}=(v_1,\dots,v_n)\in(\mathbb{F}_q^*)^{n}$ such that $\mathcal{C}_2=\Phi_{\pi,\boldsymbol{v}}(\mathcal{C}_1)$, where
	\begin{align*}
	\Phi_{\pi,\boldsymbol{v}}:\mathbb{F}_q^n\to\mathbb{F}_q^n,\quad(c_1,\ldots,c_n)\mapsto (v_1c_{\pi(1)},\ldots,v_nc_{\pi(n)}).
	\end{align*}
\end{definition}
\begin{remark}\label{es}
	It is easy to see that   $\mathcal{C}_1^2$ and $\mathcal{C}_2^2$ are equivalent when $\mathcal{C}_1$ and $\mathcal{C}_2$ are equivalent. 
\end{remark}

The definition of the punctured code is given in the following.
\begin{definition}[\cite{28}]
	For any positive integers $m$ and $n$ with $m\le n$, let $\mathcal{C}$ be a  linear code over $\mathbb{F}_{q}$ with length $n$, and $I=\{i_1,\ldots,i_m\}\subseteq \{1,\ldots,n\}$. The punctured code for  $\mathcal{C}$ over $I$ is defined as
	\begin{align*}
	\mathcal{C}_{I}=\{(c_{i_1},\ldots,c_{i_m})~|~(c_1,\ldots,c_{i_1},\ldots,c_{i_m},\ldots,c_n)\in\mathcal{C}\}. 
	\end{align*}
\end{definition}  
\begin{remark}\label{r0}	$(1)$ Recalled that $A_{\boldsymbol{\alpha}}=\{\alpha_1,\ldots,\alpha_n\}$, if $A_{\boldsymbol{\alpha}}=\mathbb{F}_q$, we always denote $\mathcal{C}_{k,n}(\boldsymbol{\alpha},\boldsymbol{v},\eta)$ to be $\mathcal{C}_{k,q}(\mathbb{F}_q,\boldsymbol{v},\eta)$.
	 
	$(2)$ Any $[1,0]$-TGRS code is equivalent to a punctured code of $\mathcal{C}_{k,q}(\mathbb{F}_q,\boldsymbol{v},\eta)$.
	
	$(3)$ If $\boldsymbol{v}_1,\boldsymbol{v}_2\in(\mathbb{F}_{q}^{*})^n$ and $\boldsymbol{\alpha}_1,\boldsymbol{\alpha}_2\in\mathbb{F}_q^{n}$ with $A_{\boldsymbol{\alpha}_1}=A_{\boldsymbol{\alpha}_2}$, then $\mathcal{C}_{k,n}(\boldsymbol{\alpha}_1,\boldsymbol{v}_1,\eta)$ and $\mathcal{C}_{k,n}(\boldsymbol{\alpha}_2,\boldsymbol{v}_2,\eta)$ are equivalent. Thus, for $0\in A_{\boldsymbol{\alpha}}$, we always set $\alpha_n=0$ without loss of generality.

\end{remark}

\subsubsection{Self-orthogonal or LCD linear codes}

The notations about LCD, self-orthogonal, self-dual or almost self-dual codes are given in the following, respectively.

For $\mathbf{a}=(a_1,\ldots,a_n)$, $\mathbf{b}=(b_1,\ldots,b_n)$ $\in\mathbb{F}_q^n$, the inner product is defined as $$\langle \mathbf{a},\mathbf{b}\rangle=\sum\limits_{i=1}^{n}a_ib_i,$$ and then 
the dual code of $\mathcal{C}$ is defined as $$\mathcal{C}^{\perp}=\{\mathbf{c}^{'}\in\mathbb{F}_q^n~|~\langle\mathbf{c}^{'},\mathbf{c}\rangle=0~ \text{for any}~\mathbf{c}\in\mathcal{C}\}.$$
%	The hull of $\mathcal{C}$ is defined as 
%	$$\mathrm{Hull}(\mathcal{C})=\dim(\mathcal{C}\cap\mathcal{C}^{\perp})$$.

If $\mathcal{C}\cap\mathcal{C}^{\perp}=0$, then $\mathcal{C}$ is LCD. 

If $\mathcal{C}\subseteq\mathcal{C}^{\perp}$, then $\mathcal{C}$ is self-orthogonal. Especially, if $\mathcal{C}=\mathcal{C}^{\perp}$, then $\mathcal{C}$ is self-dual. 

If $\mathcal{C}$ is self-orthogonal with length $n$ odd and $\dim(\mathcal{C})=\frac{n-1}{2}$, then $\mathcal{C}$ is almost self-dual.   \\

In the following lemma, we present a sufficient and necessary condition for  $\Phi_{1_{\pi},\boldsymbol{v}}(\mathcal{C}_I)$ to be self-orthogonal, where $1_{\pi}$ is the identity of permutation group with order $n$.\\
\begin{lemma}\label{lso}
	Let $n$ and $m$ be positive integers with $m\le n$,  $I=\{i_1,\ldots,i_m\}\subseteq \{1,\ldots,n\}$, and $\boldsymbol{v}=(v_1,\ldots,v_m)\in(\mathbb{F}_q^{*})^{m}$. Then for a  linear code $\mathcal{C}$ over $\mathbb{F}_{q}$ with length $n$, $\Phi_{1_{\pi},\boldsymbol{v}}(\mathcal{C}_I)$ is self-orthogonal if and only if there is some
	$\mathbf{c}\in(\mathcal{C}^2)^{\perp}$ such that for any $j\in\{1,\ldots,m\}$,
	\begin{align*} \mathrm{Supp}(\mathbf{c})=I~\text{and}~ ~c_{i_j}=v_j^2.
	\end{align*} 
\end{lemma}

{\bf Proof}. By the definition,  $\Phi_{1_{\pi},\boldsymbol{v}}(\mathcal{C}_I)$ is self-orthogonal if and only if  
\begin{align*} 
\sum\limits_{j=1}^{m}v_{j}c_{1,i_j}v_jc_{2,i_j}=0~~\text{~for~any~}\mathbf{c}_t=(c_{t,i_1},\ldots,c_{t,i_m})\in\mathcal{C}_{I}~(t=1,2),
\end{align*}namely,
\begin{align*}
\sum\limits_{j=1}^{m}v_{j}^2(c_{1,i_j}c_{2,i_j})=0~~\text{~for~any~}\mathbf{c}_t=(c_{t,i_1},\ldots,c_{t,i_m})\in\mathcal{C}_{I}~(t=1,2).
\end{align*}
Equivalently, there is some $\mathbf{c}=(c_1,\ldots,c_n)\in(\mathcal{C}^2)^{\perp}$ such that for $j=1,\ldots,m$,
 \begin{align*} \mathrm{Supp}(\mathbf{c})=I~\text{and}~ ~c_{i_j}=v_j^2.
\end{align*}$\hfill\Box$\\
%	\begin{align*}
%	\sum\limits_{j=1}^{n}c_{j}(c_{1,j}c_{2,j})=0~~\text{~for~any~}\mathbf{c}_t=(c_{t,1},\ldots,c_{t,n})\in\mathcal{C}~(t=1,2).
%	\end{align*}

A sufficient and necessary condition for the linear code $\mathcal{C}$ with  generator matrix  $G$ to be LCD is presented in the following proposition.
\begin{proposition}[Proposition $2.3$ \cite{LCD}]\label{p1}
	If $G$ is a generator matrix for the $[n, k]$ linear code $\mathcal{C}$, then $\mathcal{C}$ is an LCD code if and only if the $k \times k$ matrix $GG^\mathrm{T}$is nonsingular.
\end{proposition}

\subsection{The weight distributions of NMDS codes}It is well-known that the weight distribution of the MDS code $[n,k,n−k+1]$ over $\mathbb{F}_q$ depends only on  the values of $n$, $k$ and $q$. While for the NMDS code $[n,k,n−k]$ over $\mathbb{F}_q$,  the weight distribution depends not only on the values of $n$, $k$ and $q$, but also on the  number of the minimum weight codewords, which can be seen in the following lemma.

\begin{lemma}[Theorem $4.1$, \cite{7}]\label{NMDS}
	Let $\mathcal{C}$ be an $[n, k, n-k]$ NMDS code over $\mathbb{F}_q$ and $A_i$ $(i=0,1,\ldots,n)$ the number of codewords in $\mathcal{C}$
	with weight $i$. Then the weight distributions of $\mathcal{C}$ and $\mathcal{C}^{\perp}$ are given by
	\begin{align}\label{ANMDS}
	A_{n-k+s}=\binom{n}{k-s}\sum\limits_{j=0}^{s-1}(-1)^j\binom{n-k+s}{j}(q^{s-j}-1)\!+\!(-1)^s\binom{k}{s}A_{n-k}\quad( s=1,\ldots,k),
	\end{align}
	and
	\begin{align}\label{DNMDS}
	A_{k+s}^{\perp}=\binom{n}{k+s}\sum\limits_{j=0}^{s-1}(-1)^j\binom{k+s}{j}(q^{s-j}-1)\!+\!(-1)^s\binom{k}{s}A_{k}^{\perp}\quad(s=1,\dots,n-k).
	\end{align}
	Furthermore, $$A_{n-k}=A_{k}^{\perp}.$$	
\end{lemma}
\begin{remark}
	Let $\mathcal{C}$ be an $[n, k, n-k+1]$ MDS code, then $A_{n-k}=A_{k}^{\perp}=0$, and the weight distributions for $\mathcal{C}$ and $\mathcal{C}^{\perp}$ are given in $(\ref{ANMDS})$ and $(\ref{DNMDS})$, respectively.
\end{remark}
%%%%%%%%%%%%%%%%%%%%%%%%%%%%%%%%%%%%%%%%%%%%%%%%%%%%%%%%
%%%%%%%%%%%%%%%%%%%%%%%%%%%%%%%%%%%%%%%%%%%%%%%%%%%%%%%%
%%%%%%%%%%%%%%%%%%%%%%%%%%%%%%%%%%%%%%%%%%%%%%%%%%%%%%%%
%%%%%%%%%%%%%%%%%%%%%%%%%%%%%%%%%%%%%%%%%%%%%%%%%%%%%%%%
%%%%%%%%%%%%%%%%%%%%%%%%%%%%%%%%%%%%%%%%%%%%%%%%%%%%%%%%
\section{The properties for the $[1,0]$-TGRS code}
In this section, we give the dual code, the non-GRS (non-EGRS) property and the weight distribution for the $[1,0]$-TGRS code.

\subsection{The dual code of the $[1,0]$-TGRS code }
%In this subsection, we give a parity check matrix for $\mathcal{C}_{k,n}(\boldsymbol{\alpha},\boldsymbol{v},\eta)$, it is different from Lemma \ref{DH} that the condition $\prod\limits_{i=1}^{n}\alpha_i\neq 0$ is not need.\\

The following lemma is useful for calculating the parity check matrix of $\mathcal{C}_{k,n}(\boldsymbol{\alpha},\boldsymbol{v},\eta)$.
 \begin{lemma}\label{lu}
	For any $m\in\mathbb{N}^{+}$ and $A\subseteq\mathbb{F}_q$ with $|A|>2$, let
	$$L_A(m)=\sum\limits_{\alpha\in A}\alpha^{m}\prod\limits_{\beta\in\mathbb{F}_q\backslash A}(\alpha-\beta),$$ then
	\begin{align}\label{lq}
	L_A(m)=\begin{cases}
	0,\quad&\text{if}~0\le m\le |A|-2;\\
	-1,\quad&\text{if}~m=|A|-1.
	\end{cases}
	\end{align}
\end{lemma}

{\bf Proof.} For any $l\in\mathbb{Z}^{+}$, it is well-known that 
\begin{align}\label{lq1}
\sum\limits_{\gamma\in \mathbb{F}_q}\gamma^{l}=\begin{cases}
-1,\quad&\text{if}~(q-1)\mid l;\\
0,\quad&\text{otherwise}.
\end{cases}
\end{align}
Then by $(\ref{lq1})$ and \begin{align*}
\prod\limits_{\beta\in\mathbb{F}_q\backslash A}(\alpha-\beta)=\alpha^{q-|A|}-\sum\limits_{\beta\in\mathbb{F}_q\backslash A}\beta\alpha^{q-|A|-1}+\cdots+(-1)^{q-|A|}\prod\limits_{\beta\in\mathbb{F}_q\backslash A}\beta,
\end{align*} we have
\begin{align*}	\begin{aligned}
L_A(m)=&\sum\limits_{\alpha\in A}\alpha^{m}\prod\limits_{\beta\in\mathbb{F}_q\backslash A}(\alpha-\beta)\\
=&\sum\limits_{\alpha\in \mathbb{F}_{q}}\Big(\alpha^{q-|A|+m}-\sum\limits_{\beta\in\mathbb{F}_q\backslash A}\beta\alpha^{q-|A|-1+m}+\cdots+(-1)^{q-|A|}\prod\limits_{\beta\in\mathbb{F}_q\backslash A}\beta\alpha^m\Big)\\
=&\sum\limits_{\alpha\in \mathbb{F}_{q}}\alpha^{q-|A|+m}-\sum\limits_{\beta\in\mathbb{F}_q\backslash A}\beta\sum\limits_{\alpha\in \mathbb{F}_{q}}\alpha^{q-|A|-1+m}+\cdots+(-1)^{q-|A|}\prod\limits_{\beta\in\mathbb{F}_q\backslash A}\beta\sum\limits_{\alpha\in \mathbb{F}_{q}}\alpha^m\\
=&\begin{cases}
0,\quad&\text{if}~0\le m\le |A|-2;\\
-1,\quad&\text{if}~m=|A|-1.
\end{cases}
\end{aligned}\end{align*}
 $\hfill\Box$\\

Now, the parity check matrix of the $[1,0]$-TGRS  code is given in the following theorem.
	\begin{theorem}\label{dt}Let $P_{\boldsymbol{\alpha},j}=(-1)^{n}\prod\limits_{\alpha\in A_{\boldsymbol{\alpha}}\backslash\{\alpha_j\}}\alpha$ and $u_j=-\prod\limits_{i=1, i\neq j}^{n}(\alpha_j-\alpha_i)^{-1}$ for any $j=1,\ldots,n$.
	Then 	
	$\mathcal{C}_{k,n}(\boldsymbol{\alpha},\boldsymbol{v},\eta)$ has the parity check matrix
{\small	\begin{align*}
H_{n-k}=\left(\begin{matrix}
&\frac{u_1}{v_1}~&\frac{u_2}{v_2}~&\ldots~&\frac{u_n}{v_n}~\\
&\frac{u_1}{v_1}\alpha_1~&\frac{u_2}{v_2}\alpha_2~&\ldots~&\frac{u_n}{v_n}\alpha_n~\\
&\vdots~&\vdots&~&\vdots~\\
&\frac{u_1}{v_1}\alpha_{1}^{n-k-2}~&\frac{u_2}{v_2}\alpha_{2}^{n-k-2}~&\cdots~&\frac{u_n}{v_n}\alpha_{n}^{n-k-2}~\\
&\frac{u_1}{v_1}\big(\alpha_{1}^{n-k-1}+\eta P_{\boldsymbol{\alpha},1}\big)~&\frac{u_2}{v_2}\big(\alpha_{2}^{n-k-1}+\eta P_{\boldsymbol{\alpha},2}\big)~&\cdots~&\frac{u_n}{v_n}\big(\alpha_{n}^{n-k-1}+\eta P_{\boldsymbol{\alpha},n}\big)~\\
\end{matrix}\right),
\end{align*}}
	Namely, {\small
	\begin{align*}
	&\mathcal{C}_{k,n}^{\perp}(\boldsymbol{\alpha},\boldsymbol{v},\eta)\\
	=&\Big\{\Big(\frac{u_1}{v_1}\big(g({\alpha_{1}})+g_{n-k-1}\eta P_{\boldsymbol{\alpha},1}\big),\ldots,\frac{u_n}{v_n}\big(g(\alpha_{n})+g_{n-k-1}\eta P_{\boldsymbol{\alpha},n}\big)\Big)~\Big|~\deg g(x)\le n-k-1\Big\},
	\end{align*}}
	where  $g_{n-k-1}$ is the coefficient of $x^{n-k-1}$ in $g(x)$.
	\end{theorem}

{\bf Proof}. For any distinct elements $\alpha_{i_1},\ldots,\alpha_{i_{n-k}},\alpha_{i_{n+1-k}}\in A_{\boldsymbol{\alpha}}$ with $\prod\limits_{j=1}^{n+1-k}\alpha_{i_j}\neq0$, by extracting columns $i_1$, $\ldots$, $i_{n + 1 − k}$ of $H$ and then right multiplying it by
$\left(\begin{matrix}
	&\alpha_{i_1}~&~&(0)~\\
	&~&\ddots&~\\
	&(0)&~&\alpha_{i_{n+1-k}}~
\end{matrix}\right),$
we can get \begin{align*}
\tilde{H}=\left(\begin{matrix}
&\alpha_{i_1}~&\alpha_{i_2}~&\ldots~&\alpha_{i_{n+1-k}}~\\
&\alpha_{i_1}^2~&\alpha_{i_2}^2~&\ldots~&\alpha_{i_{n+1-k}}^2~\\
&\vdots~&\vdots&~&\vdots~\\
&\alpha_{i_1}^{n-k-1}~&\alpha_{i_2}^{n-k-1}~&\ldots~&\alpha_{i_{n+1-k}}^{n-k-1}~\\
&\alpha_{i_1}^{n-k}+\eta P_{\boldsymbol{\alpha}}~&\alpha_{i_2}^{n-k}+\eta P_{\boldsymbol{\alpha}}~&\cdots~&\alpha_{i_{n+1-k}}^{n-k}+\eta P_{\boldsymbol{\alpha}}~\\
\end{matrix}\right),
\end{align*}where $P_{\boldsymbol{\alpha}}=\prod\limits_{\alpha\in A_{\boldsymbol{\alpha}}}\alpha$.
Thus\begin{align*}
	\mathrm{Rank} (H_{n-k})\ge	\mathrm{Rank} ( \tilde{H})=n-k,
\end{align*}
which leads
  \begin{align}\label{nk}
\mathrm{Rank} (H_{n-k})=n-k.
\end{align}
For the generator matrix $G_k$ given in (\ref{G1}), let
 $G_k=\left(\begin{matrix}&\!\!\boldsymbol{g}_0~\\
&\!\!\vdots~\\
&\!\!\boldsymbol{g}_{k-1}\end{matrix}\right)$
 and
$H_{n-k}=\left(\begin{matrix}	&\!\!\!\!\boldsymbol{h}_0\\
&\!\!\!\!\vdots~\\
&\!\!\!\!\boldsymbol{h}_{n-k-1}\end{matrix}\right).$
By $(\ref{nk})$,  it is enough to prove that $G_{k}H_{n-k}^{\mathrm{T}}=\mathbf{0}_{k\times k},$ i.e., for any $s\in\{0,\ldots,k-1\}$ and $l\in\{0,\ldots,n-k-1\}$,
	\begin{align}\label{gh}
	\langle\boldsymbol{g}_s,\boldsymbol{h}_l\rangle=0.
	\end{align}
In fact, by Lemma \ref{lu} and for any $j\in\{1,\ldots,n-1\}$, $$u_j=-\prod\limits_{i=1, i\neq j}^{n}(\alpha_j-\alpha_i)^{-1}=\prod\limits_{\beta\in\mathbb{F}_q\backslash A_{\boldsymbol{\alpha}}}(\alpha_j-\beta),$$ we can obtain $(\ref{gh})$ by the following four cases.

{\bf Case $1$}. For $s=0$ and  $l\in\{0,\ldots,n-k-2\}$, we have $k+l\le n-2$, thus
\begin{align*}
\langle\boldsymbol{g}_s,\boldsymbol{h}_l\rangle
=&\sum\limits_{i=1}^{n}u_i(1+\eta\alpha_i^{k})\alpha_i^{l}=\sum\limits_{\alpha\in A_{\boldsymbol{\alpha}}}(\alpha^{l}+\eta\alpha^{k+l})\prod\limits_{\beta\in\mathbb{F}_q\backslash A_{\boldsymbol{\alpha}}}(\alpha-\beta)=0.
\end{align*}

{\bf Case $2$}. For $s\in\{1,\ldots,k-1\}$ and $l\in\{0,\ldots,n-k-2\}$, we have $1\le s+l\le n-3$, thus
\begin{align*}
	\langle\boldsymbol{g}_s,\boldsymbol{h}_l\rangle
	=&\sum\limits_{i=1}^{n}u_i\alpha_i^{s}\alpha_i^{l}=\sum\limits_{\alpha\in A_{\boldsymbol{\alpha}}}\alpha^{s+l}\prod\limits_{\beta\in\mathbb{F}_q\backslash A_{\boldsymbol{\alpha}}}(\alpha-\beta)
	=0.
\end{align*}

{\bf Case $3$}. For $s=0$ and $l=n-k-1$, one has{\small
\begin{align*}
\langle\boldsymbol{g}_s,\boldsymbol{h}_l\rangle
=&\sum\limits_{i=1}^{n}u_i(1+\eta\alpha_i^{k})\big(\alpha_{i}^{n-k-1}+\eta P_{\boldsymbol{\alpha},i}\big)\\
=&\sum\limits_{\alpha\in  A_{\boldsymbol{\alpha}}}(\alpha^{n-k-1}+\eta\alpha^{n-1}+\eta^2(-1)^{n}P_{\boldsymbol{\alpha}}\alpha^{k-1})\prod\limits_{\beta\in\mathbb{F}_q\backslash A_{\boldsymbol{\alpha}}}(\alpha-\beta)+\eta \sum\limits_{i=1}^{n}P_{\boldsymbol{\alpha},i}\prod\limits_{\beta\in\mathbb{F}_q\backslash A_{\boldsymbol{\alpha}}}(\alpha_i-\beta)\\
=&-\eta+\eta \sum\limits_{i=1}^{n}P_{\boldsymbol{\alpha},i}\prod\limits_{\beta\in\mathbb{F}_q\backslash A_{\boldsymbol{\alpha}}}(\alpha_i-\beta).
%=&-\eta+(-1)^q\eta \sum\limits_{i=1}^{n}\prod\limits_{\alpha\in A_{\boldsymbol{\alpha}}\backslash\{\alpha_i\}}\alpha \prod\limits_{\beta\in\mathbb{F}_q\backslash A_{\boldsymbol{\alpha}}}\beta\\
%+\sum\limits_{\alpha\in  A_{\boldsymbol{\alpha}}}(1+\eta\alpha_i^{k})P_{\boldsymbol{\alpha},i}\prod\limits_{\beta\in\mathbb{F}_q\backslash A_{\boldsymbol{\alpha}}}(\alpha-\beta)\\	
%	=&\sum\limits_{\alpha\in  A_{\boldsymbol{\alpha}}}(\alpha^{n-k-1}+\eta\alpha^{n-1}+\eta P_{\boldsymbol{\alpha})\prod\limits_{\beta\in\mathbb{F}_q\backslash A_{\boldsymbol{\alpha}}}(\alpha-\beta)\\	
\end{align*}}
Now, basing on (\ref{lq1}), we prove $\langle\boldsymbol{g}_s,\boldsymbol{h}_l\rangle=0$ by the following two cases.\\

$(1)$ If $0\notin A_{\boldsymbol{\alpha}}$, then $P_{\boldsymbol{\alpha},i}=(-1)^n\alpha_i^{-1}P_{\boldsymbol{\alpha}}$ $(i=1,\ldots,n)$. Thus
\begin{align*}
\langle\boldsymbol{g}_s,\boldsymbol{h}_l\rangle
=&-\eta+(-1)^n\eta P_{\boldsymbol{\alpha}} \sum\limits_{\alpha\in A_{\boldsymbol{\alpha}}}{\alpha}^{-1}\prod\limits_{\beta\in\mathbb{F}_q\backslash A_{\boldsymbol{\alpha}}}(\alpha-\beta)\\
=&-\eta+(-1)^n\eta P_{\boldsymbol{\alpha}} \sum\limits_{\alpha\in \mathbb{F}_q^{*}}{\alpha}^{-1}\prod\limits_{\beta\in\mathbb{F}_q\backslash A_{\boldsymbol{\alpha}}}(\alpha-\beta)\\
=&-\eta+(-1)^n\eta P_{\boldsymbol{\alpha}} \sum\limits_{\alpha\in \mathbb{F}_q^{*}}\prod\limits_{\beta\in\mathbb{F}_q^{*}\backslash A_{\boldsymbol{\alpha}}}(\alpha-\beta)\\
	=&{-\eta+(-1)^n\eta P_{\boldsymbol{\alpha}} \sum\limits_{\alpha\in \mathbb{F}_q^{*}}\sum_{l=0}^{q-1-n}(-1)^{l}\alpha^{q-1-n-l}\sum_{I\subseteq \mathbb{F}_q\backslash A_{\boldsymbol{\alpha}},|I|=l}\prod_{\beta\in I}\beta  }\\
=&	{-\eta+(-1)^n\eta P_{\boldsymbol{\alpha}} \sum_{l=0}^{q-1-n}(-1)^{l}\sum\limits_{\alpha\in \mathbb{F}_q^{*}}\alpha^{q-1-n-l}\sum_{I\subseteq \mathbb{F}_q\backslash A_{\boldsymbol{\alpha}},|I|=l}\prod_{\beta\in I}\beta  }\\
=&-\eta+(-1)^n\eta P_{\boldsymbol{\alpha}}\left((-1)^{q-1-n}\sum\limits_{\alpha\in \mathbb{F}_q^{*}}\prod_{\beta\in \mathbb{F}_q^{*}\backslash A_{\boldsymbol{\alpha}}}\beta  \right)\\
%=&-\eta+(-1)^n\eta P_{\boldsymbol{\alpha}}\Big( (-1)^{q-n}\prod\limits_{\beta\in\mathbb{F}_q^{*}\backslash A_{\boldsymbol{\alpha}}}\beta\Big)\\
=&-\eta+(-1)^q\eta \prod\limits_{\beta\in\mathbb{F}_q^{*}}\beta\\
=&0.
\end{align*}

$(2)$ If $0\in A_{\boldsymbol{\alpha}}$, then
\begin{align*}
\langle\boldsymbol{g}_s,\boldsymbol{h}_l\rangle
=&-\eta+(-1)^n\eta \prod\limits_{\alpha\in A_{\boldsymbol{\alpha}}\backslash\{0\}}\alpha\prod\limits_{\beta\in\mathbb{F}_q\backslash A_{\boldsymbol{\alpha}}}(-\beta)\\
=&-\eta+(-1)^q\eta \prod\limits_{\beta\in\mathbb{F}_q^{*}}\beta\\
=&0.
%=&-\eta+(-1)^q\eta \sum\limits_{i=1}^{n}\prod\limits_{\alpha\in A_{\boldsymbol{\alpha}}\backslash\{\alpha_i\}}\alpha \prod\limits_{\beta\in\mathbb{F}_q\backslash A_{\boldsymbol{\alpha}}}\beta\\
%+\sum\limits_{\alpha\in  A_{\boldsymbol{\alpha}}}(1+\eta\alpha_i^{k})P_{\boldsymbol{\alpha},i}\prod\limits_{\beta\in\mathbb{F}_q\backslash A_{\boldsymbol{\alpha}}}(\alpha-\beta)\\	
%	=&\sum\limits_{\alpha\in  A_{\boldsymbol{\alpha}}}(\alpha^{n-k-1}+\eta\alpha^{n-1}+\eta P_{\boldsymbol{\alpha})\prod\limits_{\beta\in\mathbb{F}_q\backslash A_{\boldsymbol{\alpha}}}(\alpha-\beta)\\	
\end{align*}

{\bf Case $4$}. For $s\in\{1,\ldots,k-1\}$ and $l=n-k-1$, we have $s+l\le n-1$, thus
\begin{align*}
\langle\boldsymbol{g}_s,\boldsymbol{h}_l\rangle
=&\sum\limits_{i=1}^{n}u_i\alpha_i^{s}\big(\alpha_{i}^{n-k-1}+\eta P_{\boldsymbol{\alpha},i}\big)\\
=&\sum\limits_{\alpha\in A_{\boldsymbol{\alpha}}}(\alpha^{n-k-1+s}+\eta(-1)^{n}P_{\boldsymbol{\alpha}}\alpha^{s-1})\prod\limits_{\beta\in\mathbb{F}_q\backslash A_{\boldsymbol{\alpha}}}(\alpha-\beta)\\
=&0.
\end{align*}

Now by the above discussions, we complete the proof. $\hfill\Box$\\

\begin{remark}
$(1)$ The parity check matrix for the general TGRS code has been given in \cite{ZC21}, but the form  is complicated. Especially, for the case $(t,h)=(1,0)$, it is different from the parity check matrix $H_{n,k}$ in Theorem \ref{dt}.

$(2)$ In Lemma \ref{DH}, the parity check matrix $H_{n-k}^{*}$ of  $\mathcal{C}_{k,n}(\boldsymbol{\alpha},\boldsymbol{v},\eta)$ is given under the assumption $\prod_{i=1}^{n}\alpha_i\neq 0$. However, the last row 
$\boldsymbol{h}^{*}=\left(\frac{u_1}{v_1}\big(\alpha_{1}^{n-k-1}-\eta \frac{a}{\alpha_1}\big),\cdots,\frac{u_n}{v_n}\big(\alpha_{n}^{n-k-1}-\eta\frac{a}{\alpha_n}\big)\right)$ of  $H_{n-k}^{*}$  has a little mistake. In fact, let 
$\boldsymbol{c}=\left(v_1(1+\eta\alpha_1^{k}),\ldots,v_n(1+\eta\alpha_n^{k})\right)\in \mathcal{C}_{k,n}(\boldsymbol{\alpha},\boldsymbol{v},\eta)$,  if $n$ and $q$ are odd, then by Lemma \ref{lu}, we have
\begin{align*}
	\langle\boldsymbol{h}^{*},\boldsymbol{c}\rangle=-(1+(-1)^{q-n})\eta=-2\eta\neq 0.
\end{align*}
Thus $\boldsymbol{h}^{*}\notin \mathcal{C}_{k,n}^{\perp}(\boldsymbol{\alpha},\boldsymbol{v},\eta)$.
\end{remark}
 \begin{remark}\label{rdl}%If  $0\in A_{\boldsymbol{\alpha}}$, we assume $\alpha_n=0$  without loss of generality. Now 
 	By Theorem \ref{dt}, the following two assertions hold.
 	
	$(1)$ For $0\in A_{\boldsymbol{\alpha}}$,  
\begin{align*}
\mathcal{C}_{k,n}^{\perp}(\boldsymbol{\alpha},\boldsymbol{v},\eta)=\Big\{\Big(\frac{u_1}{v_1}g({\alpha_{1}}),\ldots,\frac{u_{n-1}}{v_{n-1}}g({\alpha_{n-1}}),\frac{u_ng_0+\eta g_{n-k-1}} {v_n}\Big)~\Big|~\deg g(x)\le n-k-1\Big\},
\end{align*}
where  $g_i$ is the coefficient of $x_i$ in $g(x)$ $(i=0~\text{or}~n-k-1)$.

 	$(2)$ For $0\notin A_{\boldsymbol{\alpha}}$,   $\mathcal{C}_{k,n}^{\perp}(\boldsymbol{\alpha},\boldsymbol{v},\eta)$ is a $[1,0]$-TGRS code, that is, $$\mathcal{C}_{k,n}^{\perp}(\boldsymbol{\alpha},\boldsymbol{v},\eta)=\mathcal{C}_{n-k,n}(\boldsymbol{\alpha},\boldsymbol{w},\eta^{'}),$$ where
 	$\boldsymbol{w}=(\frac{u_1}{v_1\alpha_1},\ldots,\frac{u_n}{v_n\alpha_n})$ and $\eta^{'}=(\eta P_{\boldsymbol{\alpha}})^{-1}$. 
 	
 \end{remark}
%Basing on the Theorem \ref{dt}, we have

%\begin{corollary}\label{c1}
%	$\mathcal{C}_{k,n}^{\perp}(\boldsymbol{\alpha},\boldsymbol{v},\eta)$ is MDS or AMDS.
%\end{corollary}

%{\bf Proof}. By Theorem\ref{dt}, for any codeword $\mathbf{c}_f\in\mathcal{C}_{k,n}^{\perp}(\boldsymbol{\alpha},\boldsymbol{v},\eta)$, there exist a $f(x)=\sum\limits_{i=0}^{n-k}a_ix^{i}\in\mathbb{F}_q[x]$ such that
%\begin{align*}
%	\mathbf{c}_{f}=\Big(\frac{u_1}{v_1}f(\alpha_1),\ldots,\frac{u_n}{v_n}f(\alpha_n),1+\eta(1+S_{\boldsymbol{\alpha}})\Big).
%\end{align*}
%Thus its Hamming weight %of  $\mathbf{c}_{f}$
%\begin{align}\label{w}
%	\boldsymbol{w}_{\mathbf{c}_{f}}\ge n-(n-k)=k.
%\end{align}

%Note that $\mathcal{C}_{k,n}^{\perp}(\boldsymbol{\alpha},\boldsymbol{v},\eta)$ is with length $n$ and dimension $n-k$. Now by  $(\ref{w})$, we complete the proof. $\hfill\Box$

\subsection{The Non-GRS (non-EGRS) property for $[1,0]$-TGRS codes }
In this subsection, we show that $[1,0]$-TGRS codes are not GRS or EGRS by using the Schur product.

The Schur square for the $[1,0]$-TGRS code is given in the following lemma.
\begin{lemma}\label{LSP}If $k\ge 3$, then the following two assertions hold. \\
	
	$(1)$ For $k\ge \frac{n}{2}$, $$\mathcal{C}_{k,n}^{2}(\boldsymbol{\alpha},\boldsymbol{v},\eta)=\mathbb{F}_{q}^{n}.$$
	
	$(2)$ For $3\le k< \frac{n}{2}$,
	\begin{align*}	
	\mathcal{C}_{k,n}^{2}(\boldsymbol{\alpha},\boldsymbol{v},\eta)
	=\mathcal{C}_{2k,n}(\boldsymbol{\alpha},\boldsymbol{v}^2,\eta^2).
	\end{align*}
%	$(1)$ For odd $q$, $$\mathcal{C}_{k,n}^{2}(\boldsymbol{\alpha},\boldsymbol{v},\eta)=\mathcal{C}_{2k,n}\big(\boldsymbol{\alpha},\boldsymbol{v}^2,2^{-1}\eta\big).$$
	
%	$(2)$ For even $q$,  $$\mathcal{C}_{k,n}^{2}(\boldsymbol{\alpha},\boldsymbol{v},\eta)=\Big\{(v_1^2f({\alpha_{1}}),\ldots,v_n^2f(\alpha_{n}),f_{2k})~|~f(x)=a_{2k}x^{2k}+\sum\limits_{i=0}^{2k-2}x^i\in\mathbb{F}_q[x]\Big\}.$$
%	Furthermore, if $2k\le n$, then $\mathcal{C}_{k,n}(\boldsymbol{\alpha},\boldsymbol{v},\eta)$ is non-GRS (EGRS).
\end{lemma}

{\bf Proof}. It is enough to prove that $(1)$ and $(2)$ are both true for $\boldsymbol{v}=\mathbf{1}$. In fact, by the definition of $\mathcal{C}_{k,n}(\boldsymbol{\alpha},\boldsymbol{1},\eta)$ and Remark \ref{r1}, for $k\ge3$, we have
{\begin{align*}\begin{aligned}
&\mathcal{C}_{k,n}^{2}(\boldsymbol{\alpha},\boldsymbol{1},\eta)\\
=&\left\langle \boldsymbol{\alpha}^{i+j}, \boldsymbol{\alpha}^{i}\star(\boldsymbol{1}+\eta\boldsymbol{\alpha}^{k}), (\boldsymbol{1}+\eta\boldsymbol{\alpha}^{k})^2~(i=1,\ldots,k-1;j=1,\ldots,k-1)\right\rangle\\
=&\left\langle\boldsymbol{\alpha}^{i},\boldsymbol{\alpha}+\eta\boldsymbol{\alpha}^{k+1},\boldsymbol{\alpha}^{k-1}+\eta\boldsymbol{\alpha}^{2k-1}, \boldsymbol{1}+2\eta\boldsymbol{\alpha}^{k}+\eta^2\boldsymbol{\alpha}^{2k}~(i=2,\ldots,2k-2)\right\rangle\\
=&\left\langle\boldsymbol{\alpha}^{i}, \boldsymbol{1}+\eta^2\boldsymbol{\alpha}^{2k}~(i=1,\ldots,2k-1)\right\rangle\\
=&\begin{cases}
\mathbb{F}_{q}^{n},\quad&\text{if~}2k\ge n;\\
\mathcal{C}_{2k,n}(\boldsymbol{\alpha},\boldsymbol{1},\eta^2),\quad&\text{if~} 2k< n.\\
\end{cases}
\end{aligned}
\end{align*}}$\hfill\Box$
%Now by $(\ref{ps})$, $(1)$-$(2)$ holds, and $$\dim \big(\mathcal{C}_{k,n}^{2}(\boldsymbol{\alpha},\boldsymbol{v},\eta)\big)=2k ~~\text{for}~~2k\le n.$$
%By proposition \ref{pr}, for an $[n,k]$ GRS codes $\mathcal{C}_1$, or an $[n,k]$ EGRS codes $\mathcal{C}_2$, if $2k\le n$, then
%$$\dim \mathcal{C}_1=\dim \mathcal{C}_2=2k-1.$$
%Thus $\mathcal{C}_{k,n}(\boldsymbol{\alpha},\boldsymbol{v},\eta)$ is non-GRS and non-EGRS.

\begin{theorem}\label{NL}
	For $3\le k\le n-3$,  $\mathcal{C}_{k,n}(\boldsymbol{\alpha},\boldsymbol{v},\eta)$ is not GRS or EGRS.
\end{theorem}

{\bf Proof.} We give the proof by the following two cases.

{\bf Case $1$}. If $3\le k<\frac{n}{2}$, then $2k\le n$. By Lemma \ref{LSP}, one has
 $$ \big(\mathcal{C}_{k,n}^{2}(\boldsymbol{\alpha},\boldsymbol{v},\eta)\big)=2k,$$ and then by Proposition \ref{pr} and Remark \ref{es},
 $\mathcal{C}_{k,n}(\boldsymbol{\alpha},\boldsymbol{v},\eta)$ is not GRS or EGRS.
 
 {\bf Case $2$}. If $n-3\ge k\ge \frac{n}{2}$, then $n-k\ge 3$ and $2k-n\ge 0$. Now we prove that $\dim \mathcal{C}_{k,n}^{\perp}(\boldsymbol{\alpha},\boldsymbol{v},\eta)$ is not GRS or EGRS by the following two cases.
 
 $(1)$ For $0\notin A_{\boldsymbol{\alpha}}$, by  Lemma \ref{LSP} and  Remark $\ref{rdl}$, we have
 \begin{align*}
 	\dim \Big(\big(\mathcal{C}_{k,n}^{\perp}(\boldsymbol{\alpha},\boldsymbol{v},\eta)\big)^{2}\Big)=2(n-k).
 \end{align*}
Thus by Proposition \ref{pr} and Remark \ref{es}, 
 $\mathcal{C}_{k,n}^{\perp}(\boldsymbol{\alpha},\boldsymbol{v},\eta)$ is non-GRS or non-EGRS, and then 
 $\mathcal{C}_{k,n}(\boldsymbol{\alpha},\boldsymbol{v},\eta)$ is non-GRS or non-EGRS.

 $(2)$ For $0\in A_{\boldsymbol{\alpha}}$, by Remark $\ref{rdl}$, we know that $\boldsymbol{c}_{i}\in \mathcal{C}_{k,n}^{\perp}(\boldsymbol{\alpha},\boldsymbol{v},\eta)$ $(i=1,2,3,4)$, where
{\small \begin{align*}
 &\mathbf{c}_1=\Big(\frac{u_1}{v_1},\ldots,\frac{u_{n-1}}{v_{n-1}},\frac{u_{n}}{v_{n}}\Big),\qquad\mathbf{c}_2=\big(\frac{u_1}{v_1}\alpha_1,\ldots,\frac{u_{n-1}}{v_{n-1}}\alpha_{n-1},0\big),\\
 &\mathbf{c}_3=\big(\frac{u_1}{v_1}\alpha_{1}^{n-k-2},\ldots,\frac{u_{n-1}}{v_{n-1}}\alpha_{n-1}^{n-k-2},0\big),\quad \mathbf{c}_4=\Big(\frac{u_1}{v_1}\alpha_{1}^{n-k-1},\ldots,\frac{u_{n-1}}{v_{n-1}}\alpha_{n-1}^{n-k-1},\frac{\eta}{v_{n}}\Big).
 \end{align*}}Thus{\small
 \begin{align*}
 \boldsymbol{c}=\boldsymbol{c}_1\star\boldsymbol{c}_4-\boldsymbol{c}_2\star\boldsymbol{c}_3=\Big(0,0,\ldots,0,\frac{\eta u_n}{v_{n}^2}~\Big)\in (\mathcal{C}_{k,n}^{\perp}(\boldsymbol{\alpha},\boldsymbol{v},\eta))^{2}.
 \end{align*}}
For an $[n,k]$ GRS (EGRS) code $\mathcal{C}$, by Proposition \ref{pr},  $(\mathcal{C}^{\perp})^2$ is an $[n,2(n-k)-1]$ GRS  (EGRS)  code, and then the minimun Hamming distance
\begin{align*}
d=n+1-(2(n-k)-1)=2k-n+2\ge 2.
\end{align*}
Thus  $\boldsymbol{c}\notin(\mathcal{C}^{\perp})^2$, which implies that
$\mathcal{C}_{k,n}^{\perp}(\boldsymbol{\alpha},\boldsymbol{v},\eta)$ is not GRS or EGRS, and then 
$\mathcal{C}_{k,n}(\boldsymbol{\alpha},\boldsymbol{v},\eta)$ is not GRS or EGRS.$\hfill\Box$

\begin{remark} 
	By Theorem \ref{NL}, we can obtain Corollaries $27$ and $29$ in \cite{2022} directly.
\end{remark}

\subsection{The weight distribution of the $[1,0]$-TGRS code}In this subsection, for any $b\in\mathbb{F}_{q}$ and $B\subseteq \mathbb{F}_{q}$, let
 $$M(k,b,B)=\Big\{A\subsetneq B~\Big|~|A|=k,~\prod\limits_{\alpha\in A}\alpha=b\Big\}.$$
 In the following theorem, we show that
the weight distributions of  $\mathcal{C}_{k,n}(\boldsymbol{\alpha},\boldsymbol{v},\eta)$ and $\mathcal{C}_{k,n}^{\perp}(\boldsymbol{\alpha},\boldsymbol{v},\eta)$ can be determined by
$M(k,(-1)^{k}\eta^{-1},A_{\boldsymbol{\alpha}})$ directly.
\begin{theorem}\label{tw}
Let $A_{i}$ and $A^{\perp}_{i}$ be the number of codewords with Hamming weight $i$ $(i=0,1,\ldots,n)$ in $\mathcal{C}_{k,n}(\boldsymbol{\alpha},\boldsymbol{v},\eta)$ and $\mathcal{C}_{k,n}^{\perp}(\boldsymbol{\alpha},\boldsymbol{v},\eta)$, respectively. Then
	{\small	\begin{align*}
		&A_{n-k+s}\\
		=&\begin{cases}
		(q-1)\#M(k,(-1)^{k}\eta^{-1},A_{\boldsymbol{\alpha}}),&\quad\text{if}~s=0;\\
		\binom{n}{k-s}\sum\limits_{j=0}^{s-1}(-1)^j\binom{n-k+s}{j}(q^{s-j}-1)\!+\!(-1)^s	(q-1)\binom{k}{s}\#M(k,(-1)^{k}\eta^{-1},A_{\boldsymbol{\alpha}}),&\quad\text{if}~s=1,\dots,k;
		\end{cases}
		\end{align*}}
	and	
	{\small	\begin{align*}
		&A_{k+s}^{\perp}\\
		=&\begin{cases}
		(q-1)\#M(k,(-1)^{k}\eta^{-1},A_{\boldsymbol{\alpha}}),&\quad\text{if}~s=0;\\
		\binom{n}{k+s}\sum\limits_{j=0}^{s-1}(-1)^j\binom{k+s}{j}(q^{s-j}-1)\!+\!(-1)^s(q-1)\binom{k}{s}\#M(k,(-1)^{k}\eta^{-1},A_{\boldsymbol{\alpha}}),&\quad\text{if}~s=1,\dots,n-k.
		\end{cases}
		\end{align*}}
\end{theorem}

{\bf Proof}. By Lemma \ref{SNN}, we know that $\mathcal{C}_{k,n}(\boldsymbol{\alpha},\boldsymbol{v},\eta)$  is MDS or NMDS. Then based on Lemma \ref{NMDS}, it is enough to 
determine $A_{n-k}$.

For any $$\mathbf{c}_{f}=(v_1f(\alpha_1),\ldots,v_nf(\alpha))\in\mathcal{C}_{k,n}(\boldsymbol{\alpha},\boldsymbol{v},\eta)~~(f(x)\in\mathcal{V}_{k,1,0,\eta}),$$ let $\boldsymbol{w}_{\mathbf{c}_f}$ be the Hamming weight of $\mathbf{c}_f$, then% By the definition of $\mathcal{C}_{k,n}(\boldsymbol{\alpha},\boldsymbol{v},\eta)$,  we know that
$$\boldsymbol{w}_{\mathbf{c}_f}= n-k \text{~if and only if~}  \#\{\alpha\in A_{\boldsymbol{\alpha}}~|~f(\alpha)=0\}=k.$$
Namely, there exist some $k$-element subset $A\subsetneq A_{\boldsymbol{\alpha}}$ and $\lambda\in\mathbb{F}_{q}^{*}$, such that 
\begin{align*}
f(x)=\lambda \prod\limits_{\alpha\in A}(x-\alpha)=\lambda\bigg( x^k+\sum\limits_{i=1}^{k-1}(-1)^{k-i}\Big(\sum\limits_{I\subsetneq A,|I|=k-i}\prod\limits_{\alpha\in I}\alpha\Big) x^{i}+(-1)^{k}\prod\limits_{\alpha\in  A}\alpha\bigg),
\end{align*}
which leads that \begin{align*}
A_{n-k}=&\#\bigg(V_{k,1,0,\eta}\cap\Big\{f(x)=\lambda \prod\limits_{\alpha\in\mathbb{F}_q\backslash A}(x-\alpha)~\Big|~\lambda\neq 0,~A\subsetneq A_{\boldsymbol{\alpha}}\Big\}\bigg)\\
=&(q-1)\#\Big\{A\subsetneq A_{\boldsymbol{\alpha}}~\Big|~|A|=k,~\prod\limits_{\alpha\in A}\alpha=(-1)^{k}\eta^{-1}\Big\}\\
=&(q-1)\#M(k,(-1)^{k}\eta^{-1},A_{\boldsymbol{\alpha}}).
\end{align*}
$\hfill\Box$\\
\begin{remark}From Theorem $\ref{tw}$, Lemma \ref{SNN} can be stated equivalently as follows.
	
	$(1)$ $\mathcal{C}_{k,n}(\boldsymbol{\alpha},\boldsymbol{v},\eta)$ is MDS if and only if $\#M(k,(-1)^ {k}\eta^{-1},A_{\boldsymbol{\alpha}})=0$;
	
	$(2)$ $\mathcal{C}_{k,n}(\boldsymbol{\alpha},\boldsymbol{v},\eta)$ is NMDS if and only if $\#M(k,(-1)^{k}\eta^{-1},A_{\boldsymbol{\alpha}})>0$.
\end{remark}

For $A_{\boldsymbol{\alpha}}=\mathbb{F}_q\text{~or~}\mathbb{F}_{q}^{*}$, we determine the explicit value of $\#M(k,(-1)^{k}\eta^{-1},A_{\boldsymbol{\alpha}})$ in the following lemma.

\begin{lemma}\label{M}
	For any $b\in\mathbb{F}_q^{*}$, if $\gcd(k,q-1)=1$, then
	$$\#M(k,b,\mathbb{F}_q^*)=\#M(k,b,\mathbb{F}_q)=\frac{1}{q-1}\binom{q-1}{k}.$$
\end{lemma}

{\bf Proof.} Obviously, $\#M(k,b,\mathbb{F}_q)=\#M(k,b,\mathbb{F}_q^{*})$. Now we determine $\#M(k,b,\mathbb{F}_q^{*})$.

For $\gcd(k,q-1)=1$, $x^{k}\in\mathbb{F}_q[x]$ is a permutation polynomial over $\mathbb{F}_q$, and then for any $a,b\in\mathbb{F}_q^{*}$, there exists some $\gamma\in\mathbb{F}_q^{*}$ such that
$ab^{-1}=\gamma^k.$ Thus, for any $(\beta_1,\ldots,\beta_k)\in M(k,b,\mathbb{F}_q^{*})$, we have $$(\gamma\beta_1,\ldots,\gamma\beta_k)\in M(k,a,\mathbb{F}_q^{*}).$$
It implies that 
\begin{align*}
\#M(k,b,\mathbb{F}_q^{*})\le \#M(k,a,\mathbb{F}_q^{*}).
\end{align*}
Similarly, we have
\begin{align*}
\#M(k,a,\mathbb{F}_q^{*})\le \#M(k,b,\mathbb{F}_q^{*}).
\end{align*}
Thus for any $a,b\in\mathbb{F}_q^{*}$, we can get
\begin{align}\label{M2}
\#M(k,a,\mathbb{F}_q^{*})=\#M(k,b,\mathbb{F}_q^{*}).
\end{align}
Furthermore, it is easy to see that
\begin{align}\label{M22}
\sum_{a\in\mathbb{F}_q^{*}}\#M(k,a,\mathbb{F}_q^{*})=\binom{q-1}{k}.
\end{align}
Now by $(\ref{M2})$-$(\ref{M22})$, we have $$\#M(k,b,\mathbb{F}_q^{*})=\frac{1}{q-1}\binom{q-1}{k}.$$

$\hfill\Box$\\

\begin{remark}
	By Theorem \ref{tw} and Lemma \ref{M}, for $\gcd(k,q-1)=1$, if $A_{\boldsymbol{\alpha}}=\mathbb{F}_q\text{~or~}\mathbb{F}_{q}^{*}$,  then the weight distributions of   $\mathcal{C}_{k,n}(\boldsymbol{\alpha},\boldsymbol{v},\eta)$ and $\mathcal{C}_{k,n}^{\perp}(\boldsymbol{\alpha},\boldsymbol{v},\eta)$ can be determined explicitly. 
\end{remark}

\section{Self-orthogonal or LCD $[1,0]$-TGRS codes}	
\subsection{A sufficient and necessary condition for a $[1,0]$-TGRS code to be self-orthogonal}

Based on Lemma $\ref{LSP}$ and  Theorem \ref{dt}, $\big(\mathcal{C}_{k,n}^{2}(\mathbb{F}_q,\boldsymbol{1},\eta)\big)^{\perp}$ is given in the following lemma directly, which is useful for presenting a sufficient and necessary condition for a $[1,0]$-TGRS code to be self-orthogonal.

 \begin{lemma}\label{LSPd} For $3\le k\le q-3$, let $\gamma$ be a primitive element of $\mathbb{F}_{q}$, then the following two assertions hold.\\
 	
 	$(1)$ If $k\ge \frac{q}{2}$, then $$\big(\mathcal{C}_{k,q}^{2}(\mathbb{F}_q,\boldsymbol{1},\eta)\big)^{\perp}=\{\mathbf{0}\}.$$
 	
 	$(2)$ If $3\le k<\frac{q}{2}$, then
 	 
 	 \begin{align*}	
 	  &\big(\mathcal{C}_{k,q}^{2}(\mathbb{F}_q,\boldsymbol{1},\eta)\big)^{\perp}=
    \Big\{\big(g({\gamma^{1}}),\ldots,g(\gamma^{q-1}),g_0+\eta^2g_{q-1-2k}\big)~|~\deg g(x)\le q-1-2k\Big\},
 	 \end{align*}
 	where $g_{0}$ and $g_{q-1-2k}$ are the coefficients of $1$ and $x^{q-1-2k}$ in $g(x)$, respectively.
 \end{lemma}

 Now a sufficient and necessary condition for TGRS code to be self-orthogonal is given in the following theorem.
 \begin{theorem}\label{tso1} For $3\le k< \frac{q+1}{2}$,  the following two assertions hold.
	
	$(1)$ If $0\in A_{\boldsymbol{\alpha}}$, then $\mathcal{C}_{k,n}(\boldsymbol{\alpha},\boldsymbol{v},\eta)$ is  self-orthogonal if and ony if  there exists some
	$g(x)=\sum\limits_{i=0}^{q-1-2k}g_{i}x^{i}\in\mathbb{F}_q[x]$ such that for any $j\in \{1,\ldots,n-1\}$ and $\beta\in\mathbb{F}_q^{*}\backslash A_{\boldsymbol{\alpha}}$,
	\begin{align}\label{du1}
	g_0+\eta^2g_{q-1-2k}=v_n^2,~~~~g(\alpha_j)=v_j^2,~~~~g(\beta)=0. 
	\end{align}
	
	$(2)$ If $0\notin A_{\boldsymbol{\alpha}}$, then $\mathcal{C}_{k,n}(\boldsymbol{\alpha},\boldsymbol{v},\eta)$ is  self-orthogonal if and ony if  there exists some
	$g(x)=\sum\limits_{i=0}^{q-1-2k}g_{i}x^{i}\in\mathbb{F}_q[x]$ such that for any $j\in \{1,\ldots,n\}$ and $\beta\in\mathbb{F}_q^{*}\backslash A_{\boldsymbol{\alpha}}$,
	\begin{align}\label{du2}
	g_0+\eta^2g_{q-1-2k}=0,~~~~g(\alpha_j)=v_j^2,~~~~g(\beta)=0. 
	\end{align} 
	
\end{theorem}

%By  Lemma \ref{lso}, Remark \ref{r0} and Lemma \ref{LSPd}, we have Theorem \ref{tso1} directly.

{\bf Proof.} Let $\gamma$ be a primitive element of $\mathbb{F}_{q}$, by Lemma \ref{LSPd},  for any codeword $\boldsymbol{c}\in\big(\mathcal{C}_{k,q}^{2}(\mathbb{F}_q,\boldsymbol{1},\eta)\big)^{\perp}$, we know that there exists a
 $g(x)\in\mathbb{F}_{q}[x]$ with $\deg g(x)\le q-1-2k$ such that 
\begin{align}\label{key}
	\boldsymbol{c}=\left(g(\gamma^{1}),\ldots,g(\gamma^{q-1}),g_0+\eta^2g_{q-1-2k}\right).
\end{align}
Now we prove that $(1)$ and $(2)$ are true as follows.

$(1)$ If $0\in A_{\boldsymbol{\alpha}}$, then we can assume that $\alpha_n=0$ and $\alpha_i=\gamma^{j_i}$, where $1\le i\le n-1$ and $j_i\in \{1,\ldots,q-1\}$. 
By Lemma \ref{lso}, we know that there exists a $\boldsymbol{c}=(c_1,\ldots,c_q)\in \big(\mathcal{C}_{k,q}^{2}(\mathbb{F}_q,\boldsymbol{1},\eta)\big)^{\perp}$ such that for any $i\in \{1,\ldots,n-1\}$ and $j\in \{1,\ldots,q-1\}\backslash\{j_1,\ldots,j_{n-1}\}$,
\begin{align*}
c_{q}=v_n^{2},~~~~c_{j_i}=v_i^2,~~~~c_{j}=0.
\end{align*}
Then by $(\ref{key})$, we can get $(\ref{du1})$.

$(2)$ If $0\notin A_{\boldsymbol{\alpha}}$, then we can assume that $\alpha_i=\gamma^{j_i}$ for all $1\le i\le n$ and $j_i\in \{1,\ldots,q-1\}$. 
By Lemma \ref{lso}, we know that there exists a $\boldsymbol{c}=(c_1,\ldots,c_q)\in \big(\mathcal{C}_{k,q}^{2}(\mathbb{F}_q,\boldsymbol{1},\eta)\big)^{\perp}$ such that for any $i\in \{1,\ldots,n\}$ and $j\in \{1,\ldots,q-1\}\backslash\{j_1,\ldots,j_{n}\}$,
\begin{align*}
c_{q}=0,~~~~c_{j_i}=v_i^2,~~~~c_{j}=0.
\end{align*}
Then by $(\ref{key})$, we can get $(\ref{du2})$.
$\hfill\Box$

\begin{corollary}
If $0\in A_{\boldmath{\alpha}}$, then ${\mathcal{C}}_{k,n}(\boldsymbol{\alpha},\boldsymbol{v},\eta)$ is not self-dual.
\end{corollary}

{\bf Proof.} If $0\in A_{\boldmath{\alpha}}$ and ${\mathcal{C}}_{k,n}(\boldsymbol{\alpha},\boldsymbol{v},\eta)$ is  self-orthogonal, then by $(\ref{du1})$, there exists a $g(x)\in\mathbb{F}_{q}[x]\backslash\{0\}$ with $\deg g(x)\le q-1-2k$ such that $g(\beta)=0$ for any $\beta\in\mathbb{F}_q^{*}\backslash A_{\boldsymbol{\alpha}}$.
Thus $\deg(g(x))\ge \left|\mathbb{F}_q^{*}\backslash A_{\boldsymbol{\alpha}}\right|=q-n$, which leads $q-n\le q-1-2k$, then $n\ge 2k+1$. Hence $\mathcal{C}_{k,n}(\boldsymbol{\alpha},\boldsymbol{v},\eta)$ is not self-dual.
$\hfill\Box$\\

Now we get the sufficient and necessary condition for  $\mathcal{C}_{k,n}(\boldsymbol{\alpha},\boldsymbol{v},\eta)$ to be almost self-dual or self-dual by  Corollaries \ref{tsol1}-\ref{tsol0}.

\begin{corollary}\label{tsol1}
	Let $3\le k< \frac{q}{2}$, $\eta\in\mathbb{F}_{q}^{*}$, $\boldsymbol{\alpha}=(\alpha_1,\ldots,\alpha_{2k},\alpha_{2k+1})\in \mathbb{F}_{q}^{2k+1}$ with  $\alpha_i\neq \alpha_j$ $(i\neq j)$ and $\boldsymbol{v}=(v_1,\ldots,v_{2k},v_{2k+1})\in(\mathbb{F}_{q}^{*})^{2k+1}$. If $\alpha_{2k+1}=0$, then $\mathcal{C}_{k,2k+1}(\boldsymbol{\alpha},\boldsymbol{v},\eta)$ is almost self-dual if and only if there exists some $\lambda\in\mathbb{F}_{q}^{*}$ such that for any $j\in\{1,\ldots,2k\}$,
	\begin{align*}
	\lambda\left(-\Big(\prod\limits_{\alpha\in  A_{\boldsymbol{\alpha}}\backslash\{0\}}\alpha^{-1}\Big)+\eta^2\right)=v_{2k+1}^2,~~~~ -\lambda\alpha_j^{-1} \prod\limits_{i=1,i\neq j}^{2k}(\alpha_j-\alpha_i)^{-1}=v_j^2.
	\end{align*}	
\end{corollary}

{\bf Proof.} By Theorem \ref{tso1},   $\mathcal{C}_{k,2k+1}(\boldsymbol{\alpha},\boldsymbol{v},\eta)$ is almost self-dual if and ony if  there exists some $g(x)=\sum\limits_{i=0}^{q-1-2k}g_{i}x^{i}\in\mathbb{F}_q[x]$ such that for any $j\in \{1,\ldots,2k\}$ and $\beta\in\mathbb{F}_q^{*}\backslash A_{\boldsymbol{\alpha}}$,
\begin{align}\label{adual}
g_0+\eta^2g_{q-1-2k}=v_{2k+1}^2,~~~~g(\alpha_j)=v_j^2,~~~~g(\beta)=0.
\end{align}
Note that $\left|\mathbb{F}_q^{*}\backslash A_{\boldsymbol{\alpha}}\right|=q-1-2k$, thus $
g(x)=\lambda\prod\limits_{\beta\in \mathbb{F}_{q}^{*}\backslash A_{\boldsymbol{\alpha}}}(x-\beta)$ with $\lambda\in\mathbb{F}_{q}^{*}$. Then
$(\ref{adual})$ is equivalent to
\begin{align*}
	&v_{2k+1}^{2}=\lambda\left(\Big(\prod\limits_{\beta\in  \mathbb{F}_{q}\backslash A_{\boldsymbol{\alpha}}}\beta\Big)+\eta^2\right)=\lambda\left(-\Big(\prod\limits_{\alpha\in  A_{\boldsymbol{\alpha}}\backslash\{0\}}\alpha^{-1}\Big)+\eta^2\right),\\
	&v_{j}^{2}=\lambda\prod_{\beta\in\mathbb{F}_q\backslash A_{\boldsymbol{\alpha}}}(\alpha_j-\beta)=-\lambda\prod\limits_{\alpha \in A_{\boldsymbol{\alpha}}\backslash\{\alpha_j\}}(\alpha_j-\alpha)^{-1}=-\lambda\alpha_j^{-1} \prod\limits_{i=1,i\neq j}^{2k}(\alpha_j-\alpha_i)^{-1}.
\end{align*}
$\hfill\Box$\\

By Theorem \ref{tso1}, with similar proof as that of  Corollary \ref{tsol1}, we have the following corollary.
\begin{corollary}\label{tsol0}
	Let $3\le k< \frac{q}{2}$,  $\eta\in\mathbb{F}_{q}^{*}$, $\boldsymbol{\alpha}=(\alpha_1,\ldots,\alpha_{2k})\in(\mathbb{F}_{q}^{*})^{2k}$ with $\alpha_i\neq \alpha_j$ $(i\neq j)$ and $\boldsymbol{v}=(v_1,\ldots,v_{2k})\in(\mathbb{F}_{q}^{*})^{2k}$.  Then $\mathcal{C}_{k,2k}(\boldsymbol{\alpha},\boldsymbol{v},\eta)$ is  self-dual if and only if there exists some $\lambda\in\mathbb{F}_{q}^{*}$ such that for any $j\in \{1,\ldots,2k\}$,
	\begin{align*}
	-\prod\limits_{\alpha\in A_{\boldsymbol{\alpha}}}\alpha^{-1}+\eta^2=0,~~~~\lambda\alpha_j^{-1}\prod\limits_{i=1,i\neq j}^{2k}(\alpha_j-\alpha_i)^{-1}=v_j^2.
	\end{align*}
\end{corollary}

%\begin{corollary}\label{tso12} For $3\le k< \frac{q+1}{2}$, if $\mathcal{C}_{k,n}(\boldsymbol{\alpha},\boldsymbol{v},\eta)$ is  self-orthogonal, then $\mathcal{C}_{l,n}(\boldsymbol{\alpha},\boldsymbol{v},\eta)$ is  self-orthogonal $(l=3,\ldots,k)$.	
%\end{corollary}

%\begin{corollary}\label{tso13} For $3\le k< \frac{q+1}{2}$, we have the following two asserctions.
	
%	$(1)$ If $0\notin A_{\boldsymbol{\alpha}}$ and $\mathcal{C}_{k,n}(\boldsymbol{\alpha},\boldsymbol{v},\eta)$ is  self-orthogonal, then $\mathcal{C}_{l,n}(\boldsymbol{\alpha},\boldsymbol{v}_l,\eta)$ is  self-orthogonal $(l=3,\ldots,k)$, where 
%	$\boldsymbol{v}_l=\boldsymbol{\alpha}^{k-l}\star\boldsymbol{v}$.
	
%	$(2)$ If there exist a $\alpha_{i_0}=0$ and $\mathcal{C}_{k,n}(\boldsymbol{\alpha},\boldsymbol{v},\eta)$ is  self-orthogonal, then \mathcal{C}_{k-1-l,n-1}(\overline{\boldsymbol{\alpha}},\overline{\boldsymbol{v}}_{l},\eta)$ is  self-orthogonal $(l=0,\ldots,k-4)$, where
%	$\overline{\boldsymbol{\alpha}}$ and $\overline{\boldsymbol{v}}_{l}$ are the vectors from $\boldsymbol{\alpha}$ and $\boldsymbol{v}_l$ deleted the $i_0$-th component, respectively. 
%\end{corollary}

\subsection{Constructions for the self-orthogonal $[1,0]$-TGRS (TGRS) code}
\subsubsection{The case for $q$ even }
For $q$ even, note that any element in $\mathbb{F}_{q}$ is a square element, by Corollaries \ref{tsol1}-\ref{tsol0} and Lemma \ref{SNN}, recall that 
\begin{align}
S_{k}(\boldsymbol{\alpha})=\Big\{(-1)^{k}\prod\limits_{i\in I}^{n}\alpha_i~|~I\subsetneq\{1,\ldots,n\},~|I|=k\Big\}, 
\end{align}
then we have the Theorems \ref{CDA}-\ref{CDA1} directly.

\begin{theorem}\label{CDA}
	For any even $q$ and integer $k$ with $3\le k\le \frac{q}{2}-1$, let $\boldsymbol{\alpha}=(\alpha_1,\ldots,\alpha_{2k+1})\in \mathbb{F}_{q}^{2k+1}$ with $\alpha_i\neq \alpha_j$ $(i\neq j)$. If $0\in A_{\boldsymbol{\alpha}}$,  $\eta\in\mathbb{F}_{q}^{*}\Big\backslash\Big\{\prod\limits_{\alpha\in A_{\boldsymbol{\alpha}}\backslash\{0\}}\alpha^{-\frac{q}{2}}\Big\}$ and $\boldsymbol{v}=(v_1,\ldots,v_{2k+1})\text{~ with~}$ $$v_j=\alpha_j^{-\frac{q}{2}}\prod\limits_{i=1,i\neq j}^{2k}(\alpha_j-\alpha_i)^{-\frac{q}{2}}~ (j=1,\ldots,2k),~~~ v_{2k+1}=\Big(\prod_{\alpha\in A_{\boldsymbol{\alpha}}\backslash\{0\}}\alpha^{-1}+\eta^2\Big)^{\frac{q}{2}}.$$ Then the following two assertions hold.
	
	$(1)$ If $\eta^{-1}\in \mathbb{F}_{q}^{*}\backslash S_{k}(\boldsymbol{\alpha})$, then ${\mathcal{C}}_{k,2k+1}(\boldsymbol{\alpha},\boldsymbol{v},\eta)$ is an almost self-dual MDS code. 
	
	$(2)$ If $\eta^{-1}\in S_{k}(\boldsymbol{\alpha})$, then ${\mathcal{C}}_{k,2k+1}(\boldsymbol{\alpha},\boldsymbol{v},\eta)$ is an almost  self-dual NMDS code.
	
%	$${\mathcal{C}}_{k,2k+1}(\boldsymbol{\alpha},\boldsymbol{v},\eta)$$ is almost self-dual.  
\end{theorem}

\begin{theorem}\label{CDA1}
	For any even $q$ and integer $k$ with $3\le k\le \frac{q}{2}-1$, let $\boldsymbol{\alpha}=(\alpha_1,\ldots,\alpha_{2k})\in(\mathbb{F}_{q}^{*})^{2k}$ with $\alpha_i\neq \alpha_j$ $(i\neq j)$,  $\boldsymbol{v}=(v_1,\ldots,v_{2k})\text{~ with~} v_j=\alpha_j^{-1}\prod\limits_{i=1,i\neq j}^{2k}(\alpha_j-\alpha_i)^{-\frac{q}{2}}~ (j=1,\ldots,2k)$,  and $\eta=\prod\limits_{\alpha\in A_{\boldsymbol{\alpha}}}\alpha^{-\frac{q}{2}}$. Then we  have the following two  assertions. %then ${\mathcal{C}}_{k,2k}(\boldsymbol{\alpha},\boldsymbol{v},\eta)$ is self-dual.  
	
	$(1)$ If $\eta^{-1}\in \mathbb{F}_{q}^{*}\backslash S_{k}(\boldsymbol{\alpha})$, then ${\mathcal{C}}_{k,2k}(\boldsymbol{\alpha},\boldsymbol{v},\eta)$ is a self-dual MDS code.
	
	$(2)$ If $\eta^{-1}\in S_{k}(\boldsymbol{\alpha})$, then ${\mathcal{C}}_{k,2k}(\boldsymbol{\alpha},\boldsymbol{v},\eta)$ is a self-dual NMDS code.
\end{theorem}

%\begin{theorem}\label{CD1}
%For even $q$ and $3\le k\le \frac{q-2}{2}$, let $l$ be an integer with $3\le  l\le k$ and $\{\alpha_1,\ldots,\alpha_{2k}\}\in N(2k,0,\mathbb{F}_q)$. If $\boldsymbol{\alpha}=(\alpha_1,\ldots,\alpha_{2k})$ and  $$\boldsymbol{v}=(v_1,\ldots,v_{2k})\text{~ with~} v_j=\prod\limits_{i=1,i\neq j}^{2k}(\alpha_j-\alpha_i)^{-\frac{q}{2}}~ (j=1,\ldots,2k),$$ then ${\mathcal{C}}_{l,2k}(\boldsymbol{\alpha},\boldsymbol{v},\eta)$ is self-orthogonal. 
%\end{theorem}
%\begin{theorem}\label{CND1}For even $q$ and $3\le k\le \frac{q-2}{2}$, let $l$ be an integer with $3\le  l\le k$ and $\{\alpha_1,\ldots,\alpha_{2k}\}\in N(2k,1,\mathbb{F}_q^{*})$. If $\boldsymbol{\alpha}=(\alpha_1,\ldots,\alpha_{2k})$  and $$\boldsymbol{v}=(v_1,\ldots,v_{2k})\text{~ with~} v_j=\prod\limits_{i=1,i\neq j}^{2k}(\alpha_j-\alpha_i)^{-\frac{q}{2}}~(j=1,\ldots,2k),$$ then ${\mathcal{C}}_{l,2k+1}(\boldsymbol{\alpha},\boldsymbol{\alpha}^{k-l}\star\boldsymbol{v},\eta)$ is self-orthogonal. 
%\end{theorem}

\subsubsection{The case for $q$ odd}
 Let $\chi_q$ be the quadratic character of $\mathbb{F}_q$ and $(t_1,t_2)\in\left(\mathbb{F}_q^{*}\right)^{2}$, then
  $$\sum\limits_{a\in\mathbb{F}_q}\chi_q(t_2a^2-t_1)=-\chi_q(t_2)~ ~(\text{Theorem $5.48$ \cite{R}}).$$ Thus,
$$\{x\in\mathbb{F}_{q}^{*}~|~\chi_q(t_2x^2-1)=1\}\neq\emptyset.$$
which implies that there exist some $a\in\mathbb{F}_{q}$ such that $t_2a^2-t_1$ is a square element in $\mathbb{F}_{q}$, and then we can give a construction of almost  self-dual ${\mathcal{C}}_{\frac{q-1}{2},q}(\mathbb{F}_q,\boldsymbol{v},\eta)$ in the following Theorem.
\begin{theorem}\label{q1}
	For an odd prime power $q$, let $a\in\mathbb{F}_{q}^{*}$ such that $a^2-1$ is a square element. If  $\boldsymbol{v}=(v_1,\ldots,v_{q-1},v_{q})=(\underbrace{1,\ldots,1}_{q-1},a)$ and $\eta$ is a square root of $a^2-1$, then  ${\mathcal{C}}_{\frac{q-1}{2},q}(\mathbb{F}_q,\boldsymbol{v},\eta)$ is an almost  self-dual NMDS code.\\ 
\end{theorem}

{\bf Proof.}~By calculating directly, we have
\begin{align*}
	\Big(-\Big(\prod\limits_{\alpha\in\mathbb{F}_{q}^{*}}\alpha^{-1}\Big)+\eta^2\Big)=1+a^2-1=v_{q}^2,
\end{align*}
and for $1\le j\le q-1$,
\begin{align*}
-\alpha_j^{-1} \prod\limits_{i=1,i\neq j}^{q-1}(\alpha_j-\alpha_i)^{-1}=-\prod\limits_{\alpha\in\mathbb{F}_{q}\backslash\{\alpha_j\}}(\alpha_j-\alpha)^{-1}=-\prod\limits_{\alpha\in\mathbb{F}_{q}^{*}}\alpha=1=v_j^2.
\end{align*}
Then by Corollary \ref{tsol1}, we know that ${\mathcal{C}}_{\frac{q-1}{2},q}(\mathbb{F}_q,\boldsymbol{v},\eta)$ is almost  self-dual.
Furthermore, it  follows from Remark $\ref{SNNT}$ that ${\mathcal{C}}_{\frac{q-1}{2},q}(\mathbb{F}_q,\boldsymbol{v},\eta)$ is NMDS.$\hfill\Box$\\

Note that $\chi_q(-1)=1$ if and only if  $q\equiv 1(\mathrm{mod}~4)$, by Corollary \ref{tsol0} and Lemma \ref{SNN}, we have Theorems \ref{q20}-\ref{q2}.

\begin{theorem}\label{q20}
		For an  odd prime power $q$ with $q\equiv 3(\mathrm{mod}~4)$, $\eta\in\mathbb{F}_q^{*}$ and $\boldsymbol{v}\in(\mathbb{F}_q^{*})^{q-1}$, ${\mathcal{C}}_{\frac{q-1}{2},q-1}(\mathbb{F}_q^{*},\boldsymbol{v},\eta)$ is not self-dual.
\end{theorem}

{\bf Proof.}  Suppose ${\mathcal{C}}_{\frac{q-1}{2},q-1}(\mathbb{F}_q^{*},\boldsymbol{v},\eta)$ is self-dual, by Corollary \ref{tsol0},  we have $\eta^2=\prod\limits_{\alpha\in \mathbb{F}_{q}^{*}}\alpha=-1.$
Thus $\chi_q(-1)=1$, which leads $q\equiv 1(\mathrm{mod}~4)$, this contradicts the assumption $q\equiv 3(\mathrm{mod}~4)$.

 $\hfill\Box$

\begin{theorem}\label{q2}
Let $q$ be an odd prime power  with $q\equiv 1(\mathrm{mod}~4)$ and $\gamma$ be a primitive element of $\mathbb{F}_{q}$. If $\eta$ is a square root of $−1$, then  ${\mathcal{C}}_{\frac{q-1}{2},q-1}(\mathbb{F}_q^{*},\boldsymbol{1},\eta)$ is a self-dual NMDS code. 
\end{theorem}

{\bf Proof.}By calculating directly, we have
\begin{align*}
\left(-\Big(\prod\limits_{\alpha\in\mathbb{F}_{q}^{*}}\alpha^{-1}\Big)+\eta^2\right)=1-1=0,
\end{align*}
and for $1\le j\le q-1$,
\begin{align*}
-\alpha_j^{-1} \prod\limits_{i=1,i\neq j}^{q-1}(\alpha_j-\alpha_i)^{-1}=-\prod\limits_{\alpha\in\mathbb{F}_{q}\backslash\{\alpha_j\}}(\alpha_j-\alpha)^{-1}=-\prod\limits_{\alpha\in\mathbb{F}_{q}^{*}}\alpha=1=v_j^2.
\end{align*}Then by Corollary \ref{tsol1}, we know that ${\mathcal{C}}_{\frac{q-1}{2},q-1}(\mathbb{F}_q,\boldsymbol{v},\eta)$ is self-dual.
Furthermore, it follows from Remark $\ref{SNNT}$ that ${\mathcal{C}}_{\frac{q-1}{2},q-1}(\mathbb{F}_q,\boldsymbol{v},\eta)$ is NMDS. $\hfill\Box$\\

Let $\mathrm{C}_{2}^{0,q}=\{\beta^2~|~\beta\in\mathbb{F}_q^{*}\}$, based on Theorem \ref{tso1}, we have  Theorems \ref{q12}-\ref{q13}.

\begin{theorem}\label{q12}
	For an odd prime power $q$, let $\boldsymbol{\alpha}=(\alpha_1,\ldots,\alpha_{\frac{q-1}{2}},0)$ such that $A_{\boldsymbol{\alpha}}=\mathbb{F}_q\backslash\mathrm{C}_{2}^{0,q}$. Let $a\in\mathbb{F}_{q}^{*}$ such that $-2a^2+1$ is a square element, $\eta$ be a square root of $-2a^2+1$ and $\boldsymbol{v}=(\underbrace{1,\ldots,1}_{\frac{q-1}{2}},a)$, then the following two assertions hold.
	
	$(1)$ If $\eta^{-1}\in \mathbb{F}_{q}^{*}\backslash S_{\frac{q-1}{4}}(\boldsymbol{\alpha})$, then ${\mathcal{C}}_{\frac{q-1}{4},\frac{q-1}{2}+1}(\boldsymbol{\alpha},\boldsymbol{v},\eta)$ is an almost self-dual MDS code.
	
	$(2)$ If $\eta^{-1}\in S_{\frac{q-1}{4}}(\boldsymbol{\alpha})$, then  ${\mathcal{C}}_{\frac{q-1}{4},\frac{q-1}{2}+1}(\boldsymbol{\alpha},\boldsymbol{v},\eta)$ is an almost self-dual NMDS code.

\end{theorem}

{\bf Proof.} Denote $k=\frac{q-1}{2}$, let $g(x)=-2^{-1}\prod_{a\in\mathrm{C}_{2}^{0,q}}(x-a)$, then
\begin{align*}
	g(x)=-2^{-1}\left(x^{\frac{q-1}{2}}-1\right).
\end{align*}
Note that for any $\alpha\in A_{\boldsymbol{\alpha}}\backslash\{0\}$, $\alpha$ is not a square element, which leads $\alpha^{\frac{q-1}{2}}=-1$.
Thus 
\begin{align*}
&g_0+\eta^2g_{q-1-2k}=-2^{-1}\left(-1+(-2a^2+1)\right)=a^2,\\
&g(\alpha_j)=-2^{-1}(\alpha_{j}^{\frac{q-1}{2}}-1)=1~\text{for any~}j\in \left\{1,\ldots,\frac{q-1}{2}\right\},\\
&g(\beta)=0~\text{for any~} \beta\in\mathbb{F}_{q}^{*}\backslash A_{\boldsymbol{\alpha}}.
\end{align*}
Now by  $(\ref{du1})$, we know that ${\mathcal{C}}_{\frac{q-1}{4},\frac{q-1}{2}+1}(\boldsymbol{\alpha},\boldsymbol{v},\eta)$ is almost self-dual, and then by Lemma \ref{SNN}, we get the desired results. $\hfill\Box$\\

\begin{theorem}\label{q13}
	For an odd prime power $q$, let $\boldsymbol{\alpha}=(\alpha_1,\ldots,\alpha_{\frac{q-1}{2}})$ such that $A_{\boldsymbol{\alpha}}=\mathbb{F}_q^{*}\backslash\mathrm{C}_{2}^{0,q}$. Then the following two assertions hold.
	
	$(1)$ If $\eta^{-1}\in \mathbb{F}_{q}^{*}\backslash S_{\frac{q-1}{4}}(\boldsymbol{\alpha})$, then ${\mathcal{C}}_{\frac{q-1}{4},\frac{q-1}{2}}(\boldsymbol{\alpha},\boldsymbol{1},1)$ is a self-dual MDS code.
	
	$(2)$ If $\eta^{-1}\in S_{\frac{q-1}{4}}(\boldsymbol{\alpha})$, then ${\mathcal{C}}_{\frac{q-1}{4},\frac{q-1}{2}}(\boldsymbol{\alpha},\boldsymbol{1},1)$ is a self-dual NMDS;
	\\
\end{theorem}

{\bf Proof.} By $(\ref{du2})$, with similar proof as that of Theorem $\ref{q12}$, we can give the proof of Theorem $\ref{q13}$ directly. $\hfill\Box$\\

For any positive integer $m$, it is well known that any element in $\mathbb{F}_{p^m}^{*}$ is a square element in $\mathbb{F}_{p^{2m}}$, based on this fact, for any $\boldsymbol{\alpha}=(\alpha_1,\ldots,\alpha_{2k+1})\in(\mathbb{F}_{p^{m}})^{2k+1}$, since
$\prod\limits_{\alpha\in A_{\boldsymbol{\alpha}}\backslash\{0\}}\alpha^{-1}$  and 
$\alpha_j^{-1}\prod\limits_{i=1,i\neq j}^{2k}(\alpha_j-\alpha_i)^{-1}$ $(j=1,\ldots,2k)$ are elements in $\mathbb{F}_{p^m}^{*}$, then $\prod\limits_{\alpha\in A_{\boldsymbol{\alpha}}\backslash\{0\}}\alpha^{-1}$  and 
$\alpha_j^{-1}\prod\limits_{i=1,i\neq j}^{2k}(\alpha_j-\alpha_i)^{-1}$ $(j=1,\ldots,2k)$  are square elements in $\mathbb{F}_{p^{2m}}$. Now by Lemma \ref{SNN} and Corollaries \ref{tsol1}-\ref{tsol0}, we can get Theorems \ref{PCD1}-\ref{PCD2} directly.

\begin{theorem}\label{PCD1}
	For any $m\in\mathbb{Z}^{+}$ and odd prime $p$, $3\le k\le \frac{p^{m}-1}{2}$ and $q=p^{2m}$. let $\boldsymbol{\alpha}=(\alpha_1,\ldots,\alpha_{2k},\alpha_{2k+1})\in(\mathbb{F}_{p^{m}})^{2k+1}$ with $\alpha_i\neq \alpha_j$ $(i\neq j)$ and $a_{2k+1}=0$. Let
	 $\eta\in \mathbb{F}_{p^m}^{*}\Big\backslash\Big\{\Big(\prod\limits_{\alpha\in A_{\boldsymbol{\alpha}}\backslash\{0\}}\alpha^{-1}\Big)^\frac{1}{2}\Big\}$, $\boldsymbol{v}=(v_1,\ldots,v_{2k},v_{2k+1})$, where $v_j$ is a square root of $\alpha_j^{-1}\prod\limits_{i=1,i\neq j}^{2k}(\alpha_j-\alpha_i)^{-1}$ for any $j=1,\ldots,2k$ and $v_{2k+1}$ is a square root of $-\Big(\prod\limits_{\alpha\in  A_{\boldsymbol{\alpha}}\backslash\{0\}}\alpha^{-1}\Big)+\eta^2$, then the following two assertions hold.
	
	$(1)$ If $\eta^{-1}\in \mathbb{F}_{q}^{*}\backslash S_{k}(\boldsymbol{\alpha})$, then ${\mathcal{C}}_{k,2k+1}(\boldsymbol{\alpha},\boldsymbol{v},\eta)$ is an almost self-dual MDS code.
	
	$(2)$ If $\eta^{-1}\in S_{k}(\boldsymbol{\alpha})$, then ${\mathcal{C}}_{k,2k+1}(\boldsymbol{\alpha},\boldsymbol{v},\eta)$ is an almost  self-dual NMDS code.
		
\end{theorem}
\begin{theorem}\label{PCD2}
	For any $m\in\mathbb{Z}^{+}$ and odd prime $p$, $3\le k\le \frac{p^{m}-1}{2}$ and $q=p^{2m}$, let $\boldsymbol{\alpha}=(\alpha_1,\ldots,\alpha_{2k})\in(\mathbb{F}_{p^{m}}^{*})^{2k}$ with $\alpha_i\neq \alpha_j$ $(i\neq j)$. Let $\eta$ be a square root of $\prod\limits_{\alpha\in A_{\boldsymbol{\alpha}}}\alpha^{-1}$ and $\boldsymbol{v}=(v_1,\ldots,v_{2k})$, where $v_j$ is a square root of $\alpha_j^{-1}\prod\limits_{i=1,i\neq j}^{2k}(\alpha_j-\alpha_i)^{-1}$ for any $j=1,\ldots,2k$,  then the following two assertions hold.
	
	$(1)$ If $\eta^{-1}\in \mathbb{F}_{q}^{*}\backslash S_{k}(\boldsymbol{\alpha})$, then ${\mathcal{C}}_{k,2k}(\boldsymbol{\alpha},\boldsymbol{v},\eta)$ is a  self-dual MDS code.
	
	$(2)$ If $\eta^{-1}\in S_{k}(\boldsymbol{\alpha})$, then ${\mathcal{C}}_{k,2k}(\boldsymbol{\alpha},\boldsymbol{v},\eta)$ is a self-dual NMDS code.
	
\end{theorem}

\subsection{LCD $[1,0]$-TGRS codes from self-orthogonal  $[1,0]$-TGRS codes}
In this subsection, inspired by the Carlet et al's method for constructing LCD linear codes from  self-orthogonal linear codes \cite{LCD}, we have the following
\begin{lemma}\label{LCD}
For $q>3$, let $\mathcal{C}$ be an $[n,k]$ linear code over $\mathbb{F}_{q}$, then there exists some $I=\{i_1,\ldots,i_k\}\subsetneq\{1,\ldots,n\}$ such that $\mathcal{C}_{I}=\mathbb{F}_q^{k}$. Furthermore, for any $\beta\in\mathbb{F}_q\backslash\{0,-1,1\}$, let $\boldsymbol{v}=(v_1,\ldots,v_n)$ with $v_i=1$ $(i\in I)$ and $v_i=\beta$ $(i\notin I)$, if $\mathcal{C}$ is self-orthogonal, then $\Phi_{1_\pi,\boldsymbol{v}}(\mathcal{C})$ is LCD.
\end{lemma} 

{\bf Proof}. Let $G=\left[\begin{matrix}
\boldsymbol{g}_1,\ldots,\boldsymbol{g}_n
\end{matrix}\right]$ be a  generator matrix  of $\mathcal{C}$, where $\boldsymbol{g}_i=(g_{1,i},\ldots,g_{k,i})^{T}\in\mathbb{F}_{q}^{k}$, it is obvious that there exists some  $I=\{i_1,\ldots,i_k\}\subsetneq\{1,\ldots,n\}$ such that  $\mathcal{C}_{I}=\mathbb{F}_q^{k}$ and $\mathrm{Rank}(G_{I})=k$, where $G_{I}=\left[\begin{matrix}
\boldsymbol{g}_{i_1},\ldots,\boldsymbol{g}_{i_k}
\end{matrix}\right]$. 
Denote $I\!=\!\{i_1,\ldots,i_k\}$ with $i_1\!<\!\cdots\!<\!i_k$, and $\{j_1,\ldots,j_{n-k}\}\!=\!\{1,\ldots,n\}\backslash I$ with $j_1\!<\!\cdots\!<\!j_{n-k}$, let
\begin{align*}J_{I}=\left[\begin{matrix}
\boldsymbol{g}_{i_1},\ldots,\boldsymbol{g}_{i_{k}}
\end{matrix}\right],~~ 
P_{I}=\left[\begin{matrix}
\boldsymbol{g}_{j_1},\ldots,\boldsymbol{g}_{j_{n-k}}
\end{matrix}\right],~~J=\left[\begin{matrix}
\boldsymbol{J}_{1},\ldots,\boldsymbol{J}_{n}
\end{matrix}\right],~~
P=\left[\begin{matrix}
\boldsymbol{p}_{1},\ldots,\boldsymbol{p}_{{n}}
\end{matrix}\right],\end{align*}
where
\begin{align*}
\boldsymbol{J}_{s}=\begin{cases}
\boldsymbol{g}_s,\quad &\text{if}~s\in I;\\
\boldsymbol{0},\quad &\text{if}~s\in \{1,\ldots,n\}\backslash I, 
\end{cases}\text{\quad\quad} \boldsymbol{p}_{s}=\begin{cases}
\boldsymbol{g}_s,\quad &\text{if}~s\in \{1,\ldots,n\}\backslash I;\\
\boldsymbol{0},\quad &\text{if}~s\in I.
\end{cases}
\end{align*}	
Then  $J_{I}$ is full rank, and 
\begin{align}\label{eq1}
G=J+P,~~~JP^{\mathrm{T}}=PJ^{\mathrm{T}}=\boldsymbol{0}_{k,k},~~~JJ^{\mathrm{T}}=J_{I}J_{I}^{\mathrm{T}},~~~ PP^{\mathrm{T}}=P_{I}P_{I}^{\mathrm{T}}.
\end{align}
Furthermore, if $\mathcal{C}$ is self-orthogonal, then 
\begin{align}\label{So}
GG^{T}=\boldsymbol{0}_{k,k}.
\end{align}
Now by $(\ref{eq1})$-$(\ref{So})$, one has
\begin{align}\label{JP}
J_{I}J_{I}^{\mathrm{T}}+P_{I}P_{I}^{\mathrm{T}}=\boldsymbol{0}_{k,k}.
\end{align}  
Note that $\Phi_{1_\pi,\boldsymbol{v}}(\mathcal{C})$ has a  generator matrix  $\tilde{G}=J+\beta P$. Thus by $(\ref{eq1})$ and $(\ref{JP})$, one has
\begin{align*}
\tilde{G}\tilde{G}^{T}=J_{I}J_{I}^{\mathrm{T}}+\beta^2P_{I}P_{I}^{\mathrm{T}}=(1-\beta^2)J_{I}J_{I}^{\mathrm{T}}.
\end{align*}
Since $J_{I}$ is nonsingular and $\beta\in \mathbb{F}_q\backslash\{0,1,-1\}$, we know that $\tilde{G}\tilde{G}^{T}$ is nonsingular, then by Proposition \ref{p1},  $\Phi_{1_\pi,\boldsymbol{v}}(\mathcal{C})$ is LCD.
$\hfill\Box$\\

Basing on Lemma \ref{LCD}, we have the following theorem, which is useful for constructing LCD $[1,0]$-TGRS codes from self-orthogonal  $[1,0]$-TGRS codes.
\begin{theorem}\label{LCDT}For $q>3$, let $\beta\in\mathbb{F}_q\backslash\{0,-1,1\}$ and $\boldsymbol{w}=(\underbrace{\beta,\ldots,\beta,}_{n-k} \underbrace{ 1,\ldots,1}_{k})$. If $0\in A_{\boldsymbol{\alpha}}$ and ${\mathcal{C}}_{k,n}(\boldsymbol{\alpha},\boldsymbol{v},\eta)$ is self-orthogonal, then ${\mathcal{C}}_{k,n}(\boldsymbol{\alpha},\boldsymbol{w}\star\boldsymbol{v},\eta)$ is LCD. 
\end{theorem}

{\bf Proof}. Note that ${\mathcal{C}}_{k,n}(\boldsymbol{\alpha},\boldsymbol{v},\eta)$ is a linear code with the  generator matrix $G_k$ given in $(\ref{G1})$,
by extracting columns $n+1-k,\ldots,n$ of $G_{k}$, we can get
\begin{align*}
\tilde{G}=\left(\begin{matrix}
&v_1(1+\eta\alpha_{n+1-k}^{k})~&\cdots~&v_{n-1}(1+\eta\alpha_{n-1}^{k})~&v_n~\\
&v_1\alpha_{n+1-k}~&\cdots~&v_{n-1}\alpha_{n-1}~&0~\\
&\vdots~&\cdots~&\vdots&\vdots~\\
&v_1\alpha_{n+1-k}^{k-2}~&\cdots~&v_{n-1}\alpha_{n-1}^{k-2}~&0~\\
&v_1\alpha_{n+1-k}^{k-1}~&\cdots~&v_{n-1}\alpha_{n-1}^{k-1}~&0~\\
\end{matrix}\right)_{k\times k}
\end{align*}
Since \begin{align*}\det(\tilde{G})=
	\left|\begin{matrix}
	&v_{n+1-k}\alpha_{n+1-k}~&\cdots~&v_{n-1}\alpha_{n-1}~\\
	&\vdots~&\cdots~&\vdots~\\
	&v_{n+1-k}\alpha_{n+1-k}^{k-2}~&\cdots~&v_{n-1}\alpha_{n-1}^{k-2}~\\
	&v_{n+1-k}\alpha_{n+1-k}^{k-1}~&\cdots~&v_{n-1}\alpha_{n-1}^{k-1}~\\
	\end{matrix}\right|\neq 0,\end{align*}
we know that $\tilde{G}$ is nonsingular, which implies that the last $k$ columns of $G_k$ are linearly independent over $\mathbb{F}_q$, and so \begin{align}\label{L}
{\mathcal{C}}_{k,n}(\boldsymbol{\alpha},\boldsymbol{v},\eta)_{I}=\mathbb{F}_q^{k},~\text{where}~I=\{n+1-k,\ldots,n\}.
\end{align}

So far, by $(\ref{L})$ and Lemma $\ref{LCD}$, we get the desired result. $\hfill\Box$\\
\begin{remark}
	
	$(1)$ If $0\in A_{\boldsymbol{\alpha}}$, by Theorem \ref{LCDT}, basing on those almost self-dual ${\mathcal{C}}_{k,2k+1}(\boldsymbol{\alpha},\boldsymbol{v},\eta)$ codes constructed in subsection $4.2$, we can obtain LCD ${\mathcal{C}}_{k,2k+1}(\boldsymbol{\alpha},\boldsymbol{w}\star\boldsymbol{v},\eta)$ codes explictly.
	
	$(2)$ If $0\notin A_{\boldsymbol{\alpha}}$, by Lemma \ref{LCD}, basing on those  self-dual ${\mathcal{C}}_{k,2k}(\boldsymbol{\alpha},\boldsymbol{v},\eta)$ codes constructed in subsection $4.2$, there exist LCD $[1,0]$-TGRS codes with the same parameters. However, for $G_{k}=(\boldsymbol{g}_1,\ldots,\boldsymbol{g}_{n})$,  we can not determine which $k$-columns  $\boldsymbol{g}_{i_1},\ldots,\boldsymbol{g}_{i_k}$ are linearly independent over $\mathbb{F}_q$, thus the corresponding LCD codes can not be constructed directly.
	
 \end{remark}
\section{Conclusions }

In this paper, we got the following main results.\\

$(1)$ The weight distribution of the $[1,0]$-TGRS code is determined.\\

$(2)$  A parity check  matrix for the $[1,0]$-TGRS code is given.\\

$(3)$  $[1,0]$-TGRS codes are not GRS or EGRS.\\

$(4)$ A  sufficient and necessary condition for any punctured code of the $[1,0]$-TGRS code to be self-orthogonal is presented.\\

$(5)$ Several classses of self-dual, almost self-dual, and LCD MDS or NMDS codes are constructed. \\

\qquad\\

\noindent{\large \bf Acknowledgement~} This research was supported by the National Science Foundation of China (12071321).\\
\qquad\\

%\noindent{\large \bf Declarations}\\
\noindent{\large\bf Conflict of interest~} The authors have no conflicts of interest to declare that are relevant to the content of this
article.\\

\noindent{\large\bf Data availibility~} Not applicable.\\

\noindent{\large\bf Code Availability~} Not applicable.\\

\noindent{\large\bf Ethical approval~}Not applicable.\\

\noindent{\large\bf Consent to participate~}Not applicable.\\

\noindent{\large\bf Consent for publication~}Not applicable.\\

\end{document}